\documentclass[preprint]{aastex}

\def\ni{\noindent}

\def\cm{{\rm cm}}

\def\mm{{\rm\,mm}}
\def\gm{{\rm\,g}}
\def\d{\rm\,d}

\def\AU{{\rm\, AU}}
\def\mum{\,\mu{\rm m}}
\def\K{{\rm\,K}}
\def\yr{{\rm\,yr}}

\begin{document}

\shorttitle{Passive T Tauri and HAe Disks}
\shortauthors{Chiang et al.}

\title{Spectral Energy Distributions of Passive T Tauri and
Herbig Ae Disks: Grain Mineralogy, Parameter Dependences, and
Comparison with ISO LWS Observations}

\author{E.~I.~Chiang\altaffilmark{1,2,3}, M.~K.~Joung\altaffilmark{3,4},
M.~J.~Creech-Eakman\altaffilmark{5,6}, C.~Qi\altaffilmark{6},
J.~E.~Kessler\altaffilmark{7}, G.~A.~Blake\altaffilmark{6,7}, \&
E.~F.~van~Dishoeck\altaffilmark{8}}
\altaffiltext{1}{Hubble Fellow}
\altaffiltext{2}{Institute for Advanced Study, School of Natural Sciences,
Einstein Drive, Princeton, NJ~08540, USA}
\altaffiltext{3}{Theoretical Astrophysics, California Institute of Technology
130--33, Pasadena, CA~91125, USA}
\altaffiltext{4}{Department of Astronomy, Columbia University, New York, NY
10027}
\altaffiltext{5}{Jet Propulsion Lab, MS 171--113, Pasadena, CA~91109, USA}
\altaffiltext{6}{Division of Geological and Planetary Sciences, California
Institute of Technology 150--21, Pasadena, CA~91125, USA}
\altaffiltext{7} {Division of Chemistry and Chemical Engineering, California
Institute of Technology, Pasadena, CA~91125, USA}
\altaffiltext{8}{Sterrewacht Leiden, PO Box 9513, 2300 RA, Leiden, The
Netherlands}
\email{chiang@ias.edu, moo@astro.columbia.edu, mce@huey.jpl.nasa.gov, qch,
kessler, gab@gps.caltech.edu, ewine@strw.leidenuniv.nl}

\begin{abstract}
We improve upon the radiative, hydrostatic equilibrium
models of passive circumstellar disks constructed by
Chiang \& Goldreich (1997). New features include (1)
account for a range of particle sizes, (2) employment
of laboratory-based optical constants of representative
grain materials, and (3) numerical solution of
the equations of radiative and hydrostatic equilibrium
within the original 2-layer (disk surface + disk interior) approximation.
We systematically explore how the spectral energy
distribution (SED) of a face-on disk depends on grain
size distributions, disk geometries and surface densities,
and stellar photospheric temperatures. Observed
SEDs of 3 Herbig Ae and 2 T Tauri stars, including
spectra from the Long Wavelength Spectrometer
(LWS) aboard the Infrared Space Observatory (ISO),
are fitted with our models. Silicate emission
bands from optically thin, superheated disk surface
layers appear in nearly all systems.
Water ice emission bands appear in LWS spectra
of 2 of the coolest stars. Infrared excesses
in several sources are consistent with significant
vertical settling of photospheric grains.
While this work furnishes further evidence that passive
reprocessing of starlight by flared disks adequately
explains the origin of infrared-to-millimeter wavelength
excesses of young stars, we emphasize by explicit calculations how
the SED alone does not provide sufficient information
to constrain particle sizes and disk masses uniquely.
\end{abstract}


\section{INTRODUCTION}
\label{intro}

The energetics of the outermost regions of isolated disks
surrounding T Tauri and Herbig Ae stars is dominated
by passive reprocessing
of central starlight. While many
protostellar disks are actively accreting (see, e.g.,
the review by Calvet, Hartmann, \& Strom 2000),
effects of viscous dissipation on disk spectra
manifest themselves most strongly in the immediate vicinities of
the central stars, i.e., in the steepest portions of their
gravitational potential wells. Simple scaling laws
illustrate the relative importance of external irradiation
vs. accretion luminosity. The local viscous luminosity
per unit disk area decreases as $1/a^3$, where $a$ is
the stellocentric distance. By contrast, the flux of
central stellar radiation striking the disk drops more
slowly as $(\sin \alpha) /a^2$, where $\alpha$ is the angle
at which starlight grazes the disk surface. Vertical
hydrostatic equilibrium normally ensures that
disks flare outward such that $\alpha$ is
a slowly increasing function of $a$ for $a \gg R_*$,
where $R_*$ is the stellar radius (see, e.g.,
Kenyon \& Hartmann 1987). Hence, there is always
a disk radius outside of which the energy from stellar illumination
outweighs that of midplane accretion; in the extreme case
that the central star derives its luminosity wholly from
accretion, this transition radius is roughly 1 AU. The
spectral energy distributions (SEDs) of young star/disk systems
longward of $\sim$10$\mum$ should thus closely approximate those of passively
heated disks, even when accretion is ongoing.

Hydrostatic, radiative equilibrium models of passive T Tauri disks
are derived by Chiang \& Goldreich (1997, hereafter CG97).
The passive disk divides naturally into two regions: a surface layer
that contains dust grains directly exposed to central starlight,
and a cooler interior that is encased and diffusively heated by
the surface (Calvet et al.~1991; Malbet
\& Bertout 1991; CG97; D'Alessio et al.~1998).
CG97 compute SEDs of passive disks viewed face-on and employ their
model to satisfactorily fit the flattish infrared excess
and millimeter wavelength emission of the T Tauri star GM Aur.
The optically thin, superheated surface layer is shown to be the
natural seat of silicate emission lines (see also Calvet et al.~1992).

In a second paper, Chiang \& Goldreich (1999, hereafter CG99)
compute SEDs of passive T Tauri disks viewed at arbitrary
inclinations. They point out that the spectrum of a nearly
edge-on disk is that of a class I source for which
$\nu F_{\nu}$ rises from 2 to $10\mum$ (Lada \& Wilking 1984;
Lada 1987). In general, class I sources are
best described by a combination of an inclined,
passively heated disk and a dusty bipolar outflow or partially
evacuated envelope. The fraction of class I T Tauri spectra
that represent limiting cases of simple, isolated, inclined disks is small
(D'Alessio et al.~1999) but non-zero (CG99).

This third paper in our series on passive protostellar disks
extends our work in three directions:

\begin{enumerate}

\item We refine equilibrium, 2-layer models of passive disks
by (a) accounting for a range of particle sizes,
(b) employing laboratory-based
optical constants of a suite of circumstellar
grain materials, and (c) solving numerically the equations of radiative
and hydrostatic equilibrium within our original 2-layer approximation.

\item We systematically explore how the SED of a face-on disk depends on
grain size distributions, disk geometries and surface densities,
and stellar photospheric temperatures. Physical explanations
are provided for all observed behaviors of the SED.

\item We employ our refined face-on models to fit observed SEDs
of 3 Herbig Ae (HAe) and 2 T Tauri stars. These
observations include new spectra between 43 and $195\mum$ from
the Long Wavelength Spectrometer (LWS) aboard the Infrared Space
Observatory (ISO) (Creech-Eakman, Chiang, van Dishoeck, \& Blake 2000).
The uniqueness of our fitted values of disk parameters
is assessed, and evidence for emission lines from superheated silicates
and ices is reviewed.

\end{enumerate}

The input parameters and basic equations governing our refined
standard model are detailed in \S \ref{refin}. Results, including
a systematic exploration of how the SED varies in input
parameter space, are presented in \S\ref{refinres}.
Model fits to observations
are supplied and critically examined in \S\ref{fitherbig}. There, we also
compare our results to recent modelling efforts by Miroshnichenko et
al. (1999). Finally, we summarize our findings in \S\ref{summ}.

\section{Refined Model}
\label{refin}

\subsection{Input Parameters}
\label{ipref}

\placetable{fref}
\begin{deluxetable}{cll}
\tablewidth{0pc}
\tablecaption{Input Parameters of Refined Model\label{fref}}
\tablehead{
\colhead{Symbol}      & \colhead{Meaning} &
\colhead{Standard Value}}
\startdata
$M_*$ & Stellar Mass & $0.5 M_{\odot}$ \\
$R_*$ & Stellar Radius & $2.5 R_{\odot}$ \\
$T_*$ & Stellar Effective Temperature & $4000 \K$ \\
$\Sigma_0$ & Surface Density at 1 AU & $10^3 \, \gm\, \cm^{-2}$ \\
$p$ & $-\d\log{\Sigma}/d\log{a}$ & $1.5$ \\
$a_o$ & Outer Disk Radius & $8600 R_* = 100 \AU$ \\
$H/h$ & Visible Photospheric Height / Gas Scale Height & $4.0$ \\
$q_i$ & $-\d\log{N}/d\log{r}$ in Interior & $3.5$ \\
$q_s$ & $-\d\log{N}/d\log{r}$ in Surface & $3.5$ \\
$r_{max,i}$ & Maximum Grain Radius in Interior & $1000 \mum$ \\
$r_{max,s}$ & Maximum Grain Radius in Surface & $1 \mum$ \\
$T_{sub}^{iron}$ & Iron Sublimation Temperature & $2000 \K$ \\
$T_{sub}^{sil}$ & Silicate Sublimation Temperature & $1500 \K$ \\
$T_{sub}^{ice}$ & ${\rm\,H_2O}$ Ice Sublimation Temperature & $150 \K$ \\
$M_{\rm{DISK}}$\tablenotemark{a} & Total Disk Mass (Gas + Dust) & 0.014
$M_{\odot}$ \\
\tablenotetext{a}{The total disk mass is not an explicitly
inputted parameter but is derived from $\Sigma_0$, $p$, $a_o$,
and the inner disk cut-off radius, $a_i = 2 R_*$.}
\enddata
\end{deluxetable}

\begin{figure}
\plotone{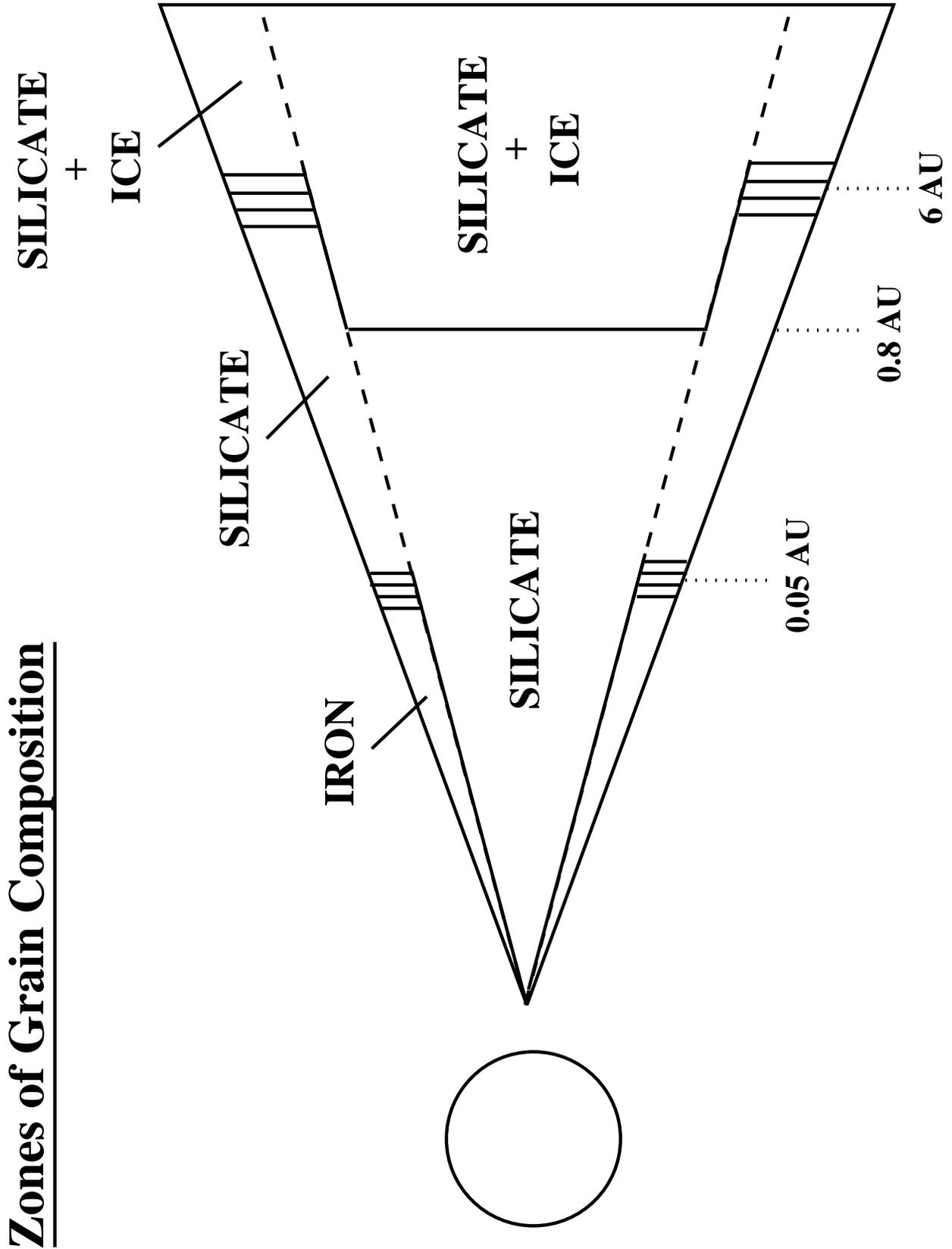}
\caption{Schematic of zones of grain compositions
(iron $\rightarrow$ amorphous olivine $\rightarrow$ amorphous olivine
mantled with water ice) for both the disk surface and interior.
The numerous vertical lines drawn in the surface indicate that
there we account for how grains of different sizes sublimate at
different stellocentric distances. Dashed lines divide the
superheated surface layers from the disk interior. Dotted lines
mark locations of condensation boundaries for the refined
standard model only.
\label{compschem}
}
\end{figure}

Table \ref{fref} lists the input parameters of our refined model.
Figure \ref{compschem} exhibits schematically the
zones of varying grain composition in both
the disk surface and disk interior. For ease of computation,
and for want of hard observational constraints on
the detailed compositions and optical properties of
circumstellar grains, we limit ourselves to considering
metallic iron (Fe, bulk density = 7.87$\gm\, \cm^{-3}$),
amorphous olivine (MgFeSiO$_4$, bulk density = 3.71$\gm\, \cm^{-3}$),
and water ice (H$_2$O, bulk density = 1$\gm\, \cm^{-3}$).
These cosmically abundant materials span a wide range
in condensation temperature (and therefore stellocentric distance),
and in the cases of silicates
and water ice, their existence is confirmed by spectroscopic
observations (see, e.g., \S\ref{isel} of this paper).
In reality, protostellar disks contain many more kinds of solid-state
materials than we have incorporated.
We have experimented with including additional grain compositions
(e.g., graphite, organics, and troilite),
but in no instance do we find our conclusions
changed qualitatively. Our goal is not to be slavishly realistic
but rather to highlight chief physical effects.

Thus, where local dust temperatures (= gas temperatures) fall below
$T_{sub}^{ice} \approx 150\K$,
the grains are taken to be spheres of amorphous olivine
mantled with water ice.
For simplicity, the thickness of the water ice mantle, $\Delta r$,
relative to the radius of the olivine core, $r$, is held constant.
Where local temperatures fall between
$T_{sub}^{ice}$ and $T_{sub}^{sil} \approx 1500\K$,
only the pure olivine cores are assumed to remain.
In innermost disk regions where local temperatures fall between
$T_{sub}^{sil}$ and $T_{sub}^{iron} \approx 2000\K$, the grains are taken to
be spheres of metallic iron.

\placefigure{compschem}

The iron or silicate cores in the disk surface
(interior) possess a power-law distribution of radii $r$ between
$r_{min}$ and $r_{max,s}$ ($r_{max,i}$):

\begin{equation}
dN \; \propto \; r^{-q_{s (i)}} \; dr \,,
\end{equation}

\ni where $dN$ is the number density of grains having radii between
$r$ and $r+dr$. Variables subscripted with ``s'' denote quantities
evaluated in the disk surface, while those subscripted with
``i'' denote quantities evaluated in the disk interior.
Our standard value of $q_i = q_s = 3.5$ places
most of the geometric surface area in the smallest grains
and most of the mass in the largest grains.
In practice, $r_{min}$ is fixed at $10^{-2} \mum$, while
$r_{max,i}$, $r_{max,s}$, $q_i$, and $q_s$ are free to vary.
Generally $r_{max,s} < r_{max,i}$
since large grains tend to settle quickly out of tenuous
surface layers (see \S3.3 of CG97). All of the cosmically abundant
iron is assumed to be locked within grains.
Following Pollack et al.~(1994), we take 50\% of the cosmically
abundant oxygen to be locked in H$_2$O ice. Values for all
cosmic abundances are obtained from Allen (2000), except
for the abundance of oxygen which is taken from Meyer, Jura, \& Cardelli
(1998).
Together, these assumptions yield a fractional
thickness, $\Delta r / r$, for the water ice mantle equal to 0.4.


Optical constants for amorphous olivine are obtained from the University
of Jena Database [\url{http://www.astro.uni-jena.de}; see also
J\"ager et al.~(1994)].
Longward of $500\mum$ where optical data for silicates
are not available, the complex refractive index ($n + ik$)
for glassy olivine is extrapolated such that
$n \, (\lambda \geq 500 \mum) = n \, (500 \mum)$ and $k \, (\lambda \geq 500
\mum) =
k \, (500 \mum) \, (\lambda / 500 \mum)^{-1}$.
Optical constants for pure crystalline H$_2$O ice are taken from the NASA ftp
site (\url{ftp : climate.gsfc.nasa.gov / pub / wiscombe / Refrac\_Index / ICE
/}; see also Warren 1984).
Though employing the constants for a cosmic mixture of
amorphous ices (${\rm H_2O}$ : $\!{\rm CH_3OH}$ : $\!{\rm CO}$ : $\!{\rm
NH_3}$;
see Hudgins et al.~1993) would be more appropriate,
we nonetheless adopt the data for pure H$_2$O ice because the latter are
available over
all wavelengths of interest, from the ultraviolet to the radio,
whereas the former are not. One consequence of using the constants for
crystalline (213--272 K) water ice as opposed to amorphous ($\sim$100 K)
ice is that spectral features due to translational lattice modes
at $45$ and $62\mum$ are slightly underestimated in width and overestimated
in amplitude (see, e.g., Hudgins et al.~1993).
Optical constants for metallic Fe are obtained from
Pollack et al.~(1994).

The inner cutoff radius of the disk, $a_i$,
is fixed at $2 R_*$. For T Tauri stellar parameters,
this radius coincides with the distance at which iron grains
in the surface layer attain their sublimation temperature.
For the hotter HAe stars, the iron condensation boundary
occurs at $a \approx 14$--$30\, R_*$.
Inside the iron condensation radius,
the disk may still be optically thick
to stellar radiation even if dust is absent.
Opacity sources include pressure-broadened
molecular lines and Rayleigh scattering
off hydrogen atoms [see the appendix of
Bell \& Lin (1994)].
For simplicity, when modelling HAe stars,
we employ a 1-layer blackbody disk that extends from the
iron condensation boundary to $a_i = 2 R_*$.
%

\subsection{Grain Absorption Efficiencies and Opacities}
\label{gaeo}

\begin{figure}
\plotone{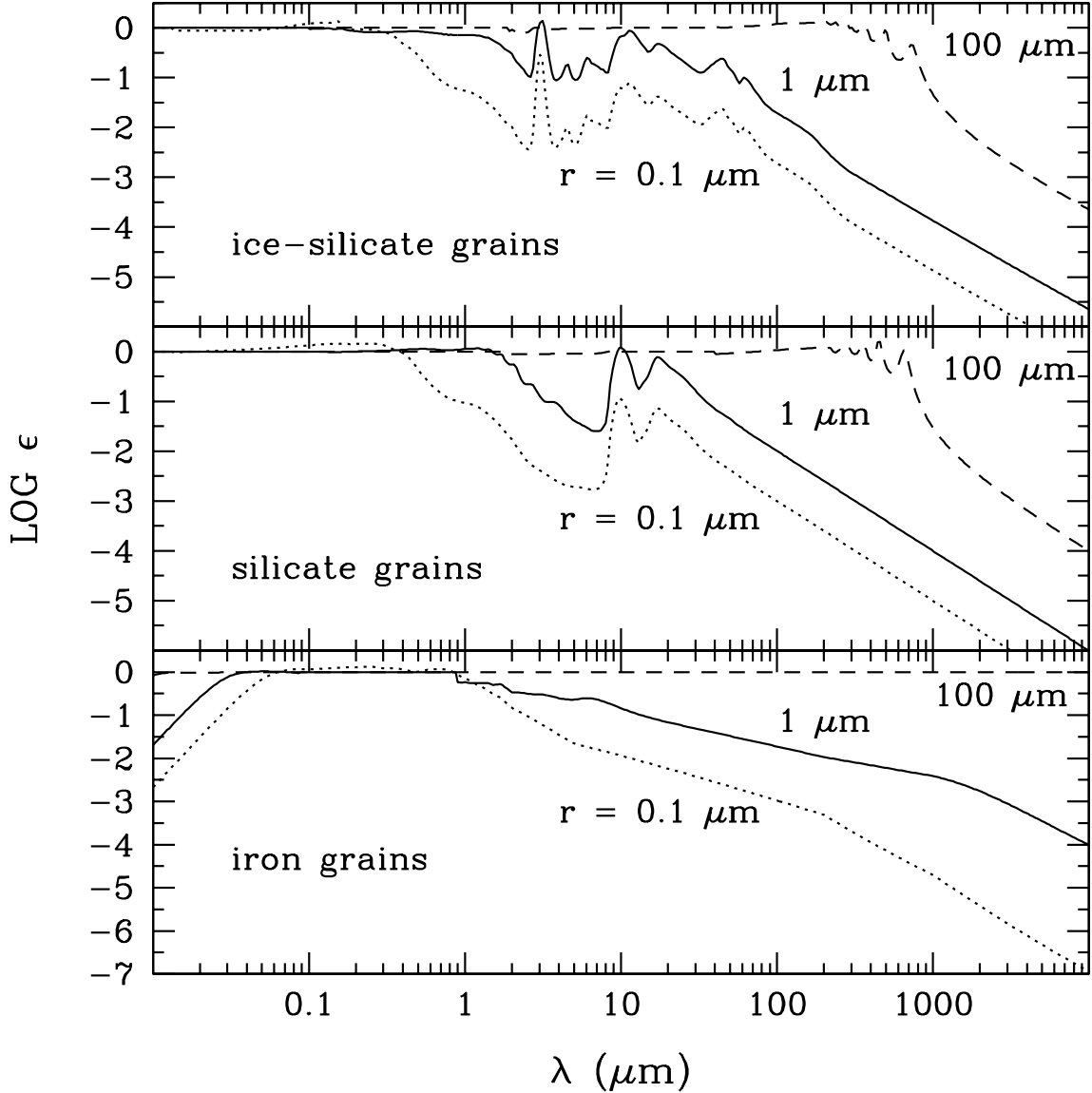}
\caption{Emissivities of ice-silicate
(H$_2$O / amorphous olivine),
silicate (amorphous olivine), and iron grains having three representative
core sizes. The thickness of the water ice mantle relative
to the radius of the olivine core is $\Delta r / r = 0.4$.
Resonant features include the
O-H stretching ($3.1$, $4.5\mum$) and H-O-H bending ($6.1\mum$)
modes in water ice; the Si-O stretching ($10\mum$) and O-Si-O bending
($18\mum$) modes in silicates; and the intermolecular translational
($45$, $62$, and $154\mum$) modes in water ice.
Oscillatory behavior near the onset of the Rayleigh limit
($2\pi r / \lambda \approx 1$) reflects ``ripple structure''
arising from our use of perfectly spherical particles.
\label{qabs}}
\end{figure}

\placefigure{qabs}

The grain emissivity, $\varepsilon (r,\lambda)$, is equal
to its absorption efficiency and is calculated using
Mie-G\"{u}ttler theory (Bohren \&
Huffman 1983; see their subroutine BHCOAT.F).\footnote{Our
grain emissivity, $\varepsilon$, equals $Q_{abs}$ in the notation
of Bohren \& Huffman (1983).}
Figure \ref{qabs} displays absorption efficiencies
for our ice-silicate, silicate, and iron spheres having three
representative sizes. The emissivity index in the Rayleigh limit,
$\beta \equiv d\ln \varepsilon / d\ln \nu$,
equals 1.64, 2.00, and 0.50 for the three compositions,
respectively.

Well-known resonances at $\lambda \lesssim 20\mum$ include the
O-H stretching ($3.1$, $4.5\mum$) and H-O-H bending ($6.1\mum$)
modes in water, and the Si-O stretching ($10\mum$) and O-Si-O bending
($18\mum$) modes in silicates.

In crystalline water ice (and crystalline silicates),
optically active modes of vibration longward of $\sim$$10\mum$
are ``intermolecular translational'' or ``intermolecular rotational.''
These involve collective movement of a molecule or
a unit cell with respect to other molecules/unit cells in the lattice.
The strengths, positions, and widths of these modes are
more sensitive to the presence of chemical impurities and to
long range order in the solid (i.e., its degree of
crystallinity, or ``allotropic state'') than those of
fundamental stretching and bending modes at shorter
wavelengths. In principle, these intermolecular modes
provide information on the annealing history of
initially amorphous, interstellar material in the
relatively high density and high temperature environments
of circumstellar disks. The intermolecular translational
modes in water ice evident in Figure \ref{qabs} are located at
$45$, $62$, and $154\mum$ (Bohren \& Huffman 1983, see
page 278; Bertie, Labb\'{e}, \& Whalley 1969, see
their Figures 4 and 11). All of these features are positioned
within the wavelength range of the Long Wavelength Spectrometer
aboard ISO. Note also the blending of the $12\mum$
intermolecular rotational band in water ice (Bohren \& Huffman 1983)
with the Si-O silicate stretching mode at $10\mum$.

Oscillatory behavior in Figure \ref{qabs} near the onset of the Rayleigh limit
($2\pi r / \lambda \approx 1$) reflects so-called ``ripple structure''
that arises from our use of perfectly spherical particles
(Bohren \& Huffman 1983); we expect real-world deviations
from sphericity to smooth out this artificial behavior.

In the disk interior, the opacity is given by

\begin{equation}
\kappa_i(\lambda) = \frac{\pi}{\rho_t} \int_{r_{min}}^{r_{max,i}} {dN \over dr}
\, r^2 \, \varepsilon (\lambda,r) \, dr \, ,
\end{equation}

\begin{figure}
\plotone{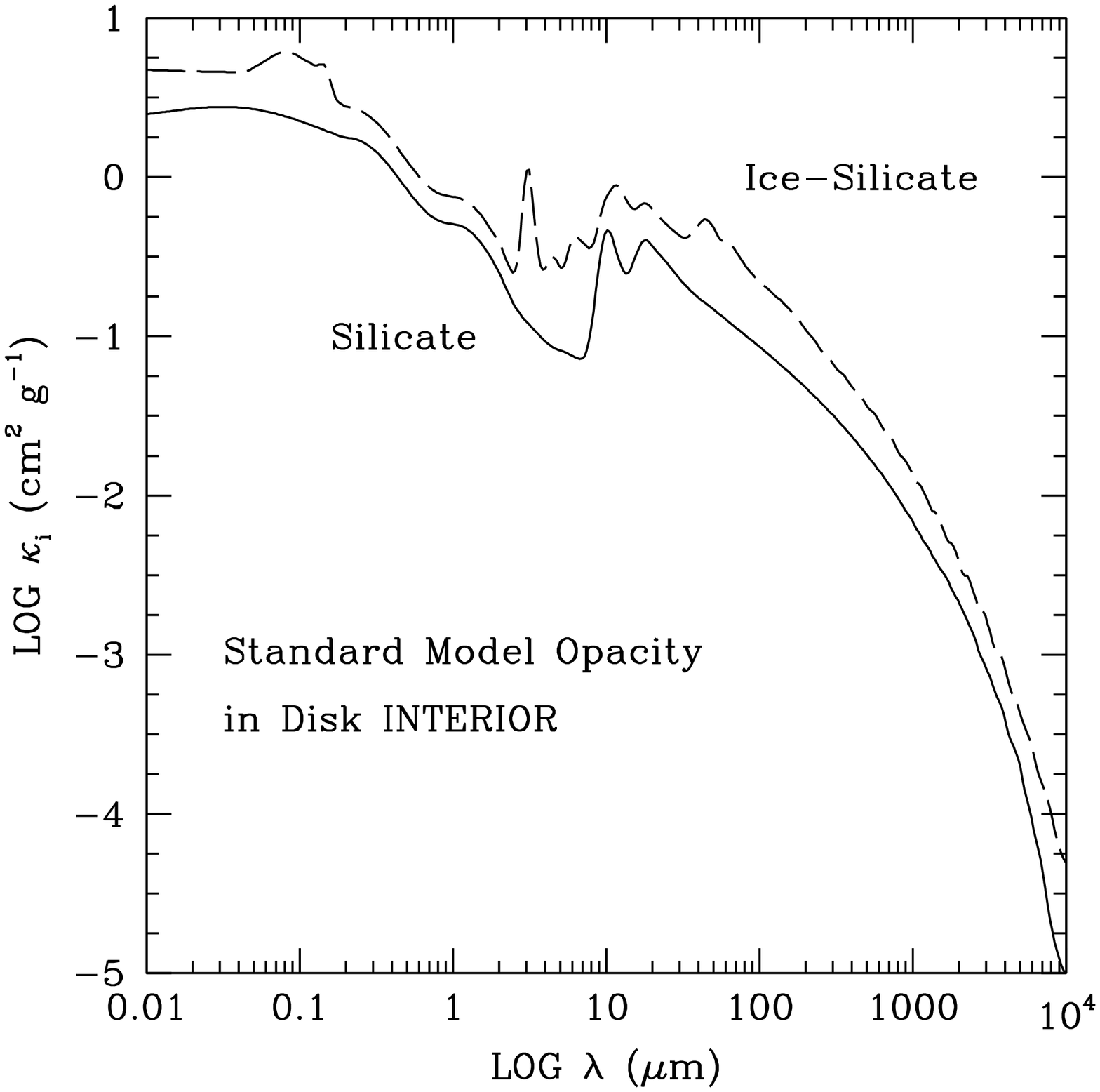}
\caption{Mass absorption coefficients for our standard model distributions
of ice-silicate and silicate particles in solar abundance gas.
The same resonant features found in Figure \protect{\ref{qabs}} are seen here.
We have smoothed these curves to suppress so-called ``ripple structure''
at $\lambda \gtrsim 1000 \mum$ that arises from the sphericity
of our particles having $r \approx 1000\mum$. Note that these curves
are sensitive to our chosen $q_i$ and $r_{max,i}$.
\label{kap}}
\end{figure}

\ni where $\rho_t$ is the total density of gas and dust.
Figure \ref{kap} displays $\kappa_i$ for our distribution of
ice-silicate and silicate particles in cosmic abundance gas.

\placefigure{kap}

\subsection{Basic Equations}
\label{beqns}

In the surface layer, dust grains are directly exposed to stellar
radiation. The grains attain an equilibrium temperature

\begin{equation}
\label{tds1}
T_{ds}(a,r)=T_* \left( \frac{R_*}{a} \right)^{1/2} \left( \frac{\phi}{4}
\frac{\langle\varepsilon (r)\rangle _{T_*}}{\langle \varepsilon (r)
\rangle_{T_{ds}}} \right)^{1/4} \, ,\
\end{equation}

\ni where $\langle\varepsilon (r) \rangle_T$ is the
emissivity of a grain of radius $r$ averaged over the Planck
function at temperature $T$, and $\phi \approx 1/2$ is the fraction of
the stellar hemisphere that is seen by the grain.

In our two-layer formalism, exactly half of the radiation reprocessed by the
surface layer escapes directly into space. The remaining half
is directed towards the disk interior. Of this latter
half, a fraction,
$1-e^{-\Sigma\langle\kappa_i\rangle _s}$, is absorbed by
the disk interior, where $\Sigma$ is the disk surface density and
$\langle\kappa_i\rangle _s$ is the opacity of the
disk interior averaged over the spectrum of radiation from the
surface.\footnote{In practice, we perform this average over a Planck function
evaluated
at the temperature of the most luminous grains in the surface.
For our assumed size distribution, these dominant grains typically
have radii $r \approx 0.5\mum$; surface grains having
$r \approx 0.2$--$1\mum$ are responsible for absorbing $\sim$50\% of
the incident stellar radiation.}
In radiative balance,

\begin{equation}
\label{ieb}
\frac{\phi}{2} \, (1- e^{-\Sigma\langle\kappa_i\rangle _s}) \,
\left({\frac{R_*}{a}}\right)^2 T_*^4 \, \sin {\alpha} = (1-
e^{-\Sigma\langle\kappa_i\rangle _i}) \, T_i^4 \, ,\
\end{equation}

\ni where $T_i$ is the temperature of the disk interior and
$\langle\kappa_i\rangle _i$ is the interior opacity averaged over
the spectrum of radiation from the disk interior.\footnote{We perform this
average over a Planck function evaluated at temperature $T_i$.}
Note that $T_i$ is the common temperature of interior
grains of all sizes which are assumed to have
thermally equilibrated with one another. Here
$\alpha$ is the angle at which stellar radiation
strikes the surface:

\begin{equation}
\label{alpha1}
\alpha \approx \arctan \left( \frac{d\ln H}{d\ln a} \frac{H}{a} \right) -
\arctan{\frac{H}{a}}+\arcsin \left( \frac{4}{3\pi}\frac{R_*}{a} \right)
\end{equation}

\ni [cf. equation (5) of CG97]. The height
of the disk photosphere, $H$, is assumed to be proportional to
the vertical gas scale height, $h$, with a fixed constant of proportionality
equal to 4 for our standard model. In reality,
when dust and gas are well-mixed in interstellar proportions,
the ratio $H/h$ decreases
slowly from $\sim$5 at 1 AU to $\sim$4 at 100 AU.
In modelling observed SEDs, we will allow $H/h$ to be a fitted
constant parameter. In hydrostatic equilibrium,

\begin{equation}
\label{ha}
\frac{H}{a} = \frac{H}{h} \frac{h}{a} = \frac{H}{h} \sqrt \frac{T_i}{T_c} \sqrt
\frac{a}{R_*} \, .\
\end{equation}

\ni where $T_c \equiv GM_*\mu_g/kR_*$ and $\mu_g = 3 \times 10^{-24} \gm$.

Equations (\ref{ieb}) and (\ref{ha}) are two equations for the
two unknown functions, $H(a)$ and $T_i(a)$.
Substitution of (\ref{ha}) into (\ref{ieb}) yields
an algebraic equation for $T_i$ and the slowly-varying
flaring index, $\gamma \equiv d\ln H / d\ln a$.
We solve this equation for $T_i(a)$ and $\gamma (a)$
numerically on a logarithmic grid
in stellocentric distance. Our procedure is
described in detail in the Appendix.

The SED of the disk equals the sum of emission from the disk
interior,

\begin{equation}
4\pi d^2\lambda F_{\lambda, i} = 8\pi^2 \lambda \int_{a_i}^{a_o} \,
B_{\lambda}(T_i) \, (1 - e^{-\Sigma \kappa_i}) \, a \, da \, ,
\end{equation}

\ni and from the surface layers (above and below the disk midplane),

\begin{equation}
4\pi d^2\lambda F_{\lambda, s} = 8\pi^2 \lambda \, (1 + e^{-\Sigma \kappa_i})
\, \int_{a_i}^{a_o} \, S_{\lambda} \, (1 - e^{-\tau_s}) \, a\, da \, ,
\end{equation}

\ni where $d$ is the distance to the source. The source function
in the surface,
$S_{\lambda}$, is the Planck function
averaged over the ensemble of effective grain cross-sections in
the surface layer:

\begin{equation}
\label{slambda}
S_{\lambda} = \frac{2 \int_{r_{min}}^{r_{max,s}} B_{\lambda} (T_{ds}) \, {dN
\over dr} \, r^2 \, \varepsilon (\lambda, r) \, dr}{\int_{r_{min}}^{r_{max,s}}
{dN \over dr} \, r^2 \, \varepsilon (\lambda, r) \, dr} \, .
\end{equation}

\ni In equation (\ref{slambda}), the factor of 2 is
inserted so that exactly half of the incident
radiation is reprocessed by the surface layer.
The normal optical depth of the surface layer, $\tau_{\lambda}$,
is similarly averaged:

\begin{equation}
\tau_s (\lambda, a) = \frac{\int_{r_{min}}^{r_{max,s}} {dN \over dr} \, r^2 \,
\varepsilon (\lambda, r) \, dr}{\int_{r_{min}}^{r_{max,s}} {dN \over dr} \, r^2
\, \langle \varepsilon (r) \rangle _{T_*} \, dr} \, \sin \alpha \, .
\end{equation}

\section{Results}
\label{refinres}

\subsection{Flaring Index}
\label{geomfan}

\begin{figure}
\plotone{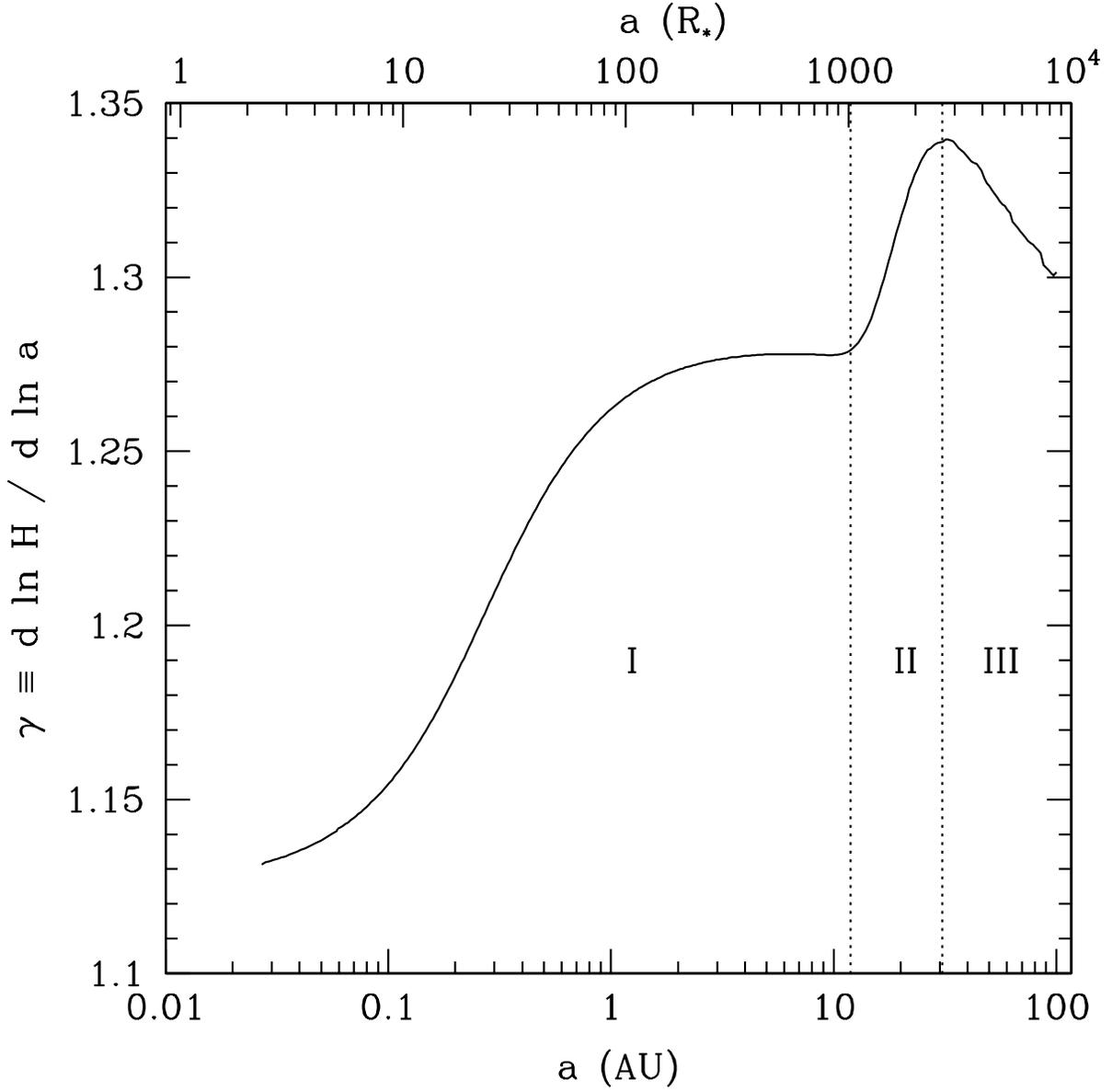}
\caption{The flaring index, $\gamma \equiv d\ln H / d\ln a$,
for our refined standard model.
As was done in CG97 (see their \S2.3.2), we divide the disk into
three annular regions depending on the optical depth of the disk
interior. In region ``I,'' the interior behaves as a blackbody;
$\gamma$ increases from its flat disk value of $1.125 \approx 9/8$
to its asymptotic flared value of $1.275 \approx 9/7$ as the
disk thickness becomes increasingly larger than the
stellar radius. In region ``II,''
the disk interior becomes optically thin to its own reprocessed
radiation; $\gamma$ increases as interior grains enhance
their temperatures to compensate for the inefficiency
with which they re-radiate. In region ``III,''
the interior is transparent to radiation from the surface
and cools quickly with increasing distance, causing $\gamma$ to decrease.
\label{gam}}
\end{figure}

Figure \ref{gam} displays the behavior of the flaring index,
$\gamma \equiv d\ln H / d\ln a$.
We conceptually divide the disk into three annular regions, as
was done in CG97 (see their \S2.3.2).
In the region marked ``I,'' the disk interior is opaque to both
its own reprocessed radiation and to radiation from the surface.
Here $\gamma$ increases from its flat disk value of $1.125 \approx 9/8$
to its asymptotic value of $1.275 \approx 9/7$ as the first
two terms on the right-hand side of equation (\ref{alpha1})
gradually dominate the last term. In region ``II,'' the disk
interior remains opaque to radiation from the surface,
but is optically thin to its own reprocessed radiation.
Here $\gamma$ steeply rises with $a$ because grains in the disk interior
equilibrate at relatively high temperatures to compensate
for the relative inefficiency with which they re-radiate
the incident energy. Finally, in region ``III,'' the interior is transparent
to radiation from the surface (i.e., $\Sigma \langle \kappa_i \rangle_s
\lesssim 1$);
the inability of the interior to absorb the incident energy
causes $\gamma$ to decrease.

\placefigure{gam}

\subsection{Disk Temperatures}
\label{dtempfan}

\begin{figure}
\plotone{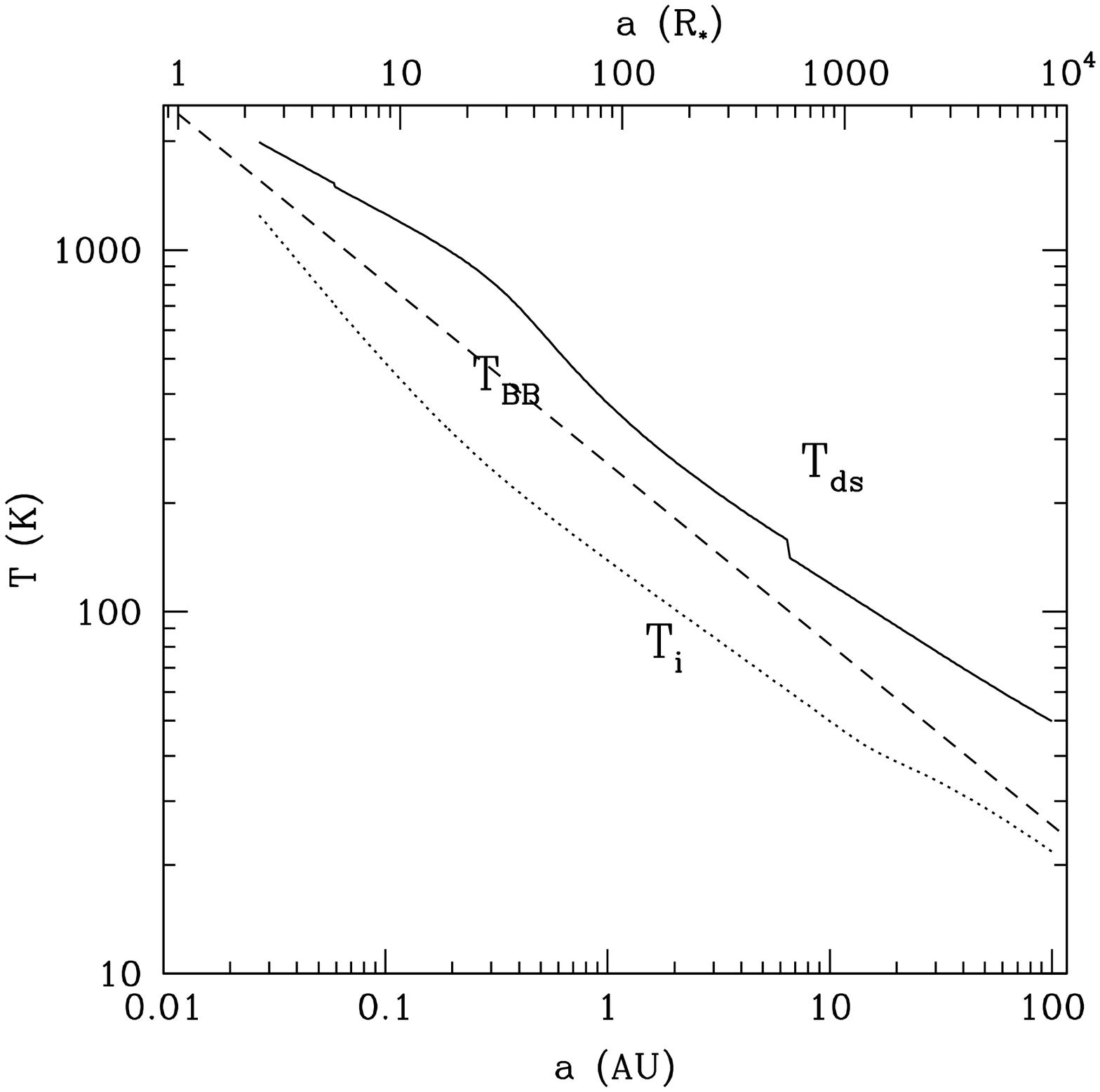}
\caption{Temperature profiles for the surface and for
the interior in our refined standard model.
The temperatures of grains in the surface layer depend
on their sizes; here, the curve marked $T_{ds}$ represents
the size bin, $r\approx 0.5\mum$, containing the most luminous grains.
The discontinuity in $T_{ds}$ at $a \approx 6\AU$ marks
the water condensation boundary in the surface, outside of which H$_2$O
ice coats silicate cores; the discontinuity in $T_{ds}$ at
$a \approx 0.05\AU$ marks the silicate condensation boundary
in the surface. For reference,
the temperature of a spherical blackbody which is naked
before half of the stellar hemisphere is shown as a dashed line.
\label{tempfancy}}
\end{figure}

\placefigure{tempfancy}

Figure \ref{tempfancy} exhibits temperature profiles for the
surface ($T_{ds}$) and for the interior ($T_i$).
The temperatures of grains in the surface layer vary
slightly with their sizes; in Figure \ref{tempfancy}, we have chosen
to plot $T_{ds}$ for the size bin containing the most luminous grains,
i.e., the logarithmic size interval that absorbs the greatest fraction of
incident stellar radiation. For our choices of grain composition and size
distribution, these dominantly absorbing grains have radii
$r \approx 0.1$--$0.7 \mum$. For reference,
we also overlay in Figure \ref{tempfancy}
the temperature of an imaginary blackbody sphere, $T_{BB}$, which
is naked before half of the stellar hemisphere.

These temperature profiles are largely similar to those found in the
simpler model presented in CG97, and help to justify
the approximations made there. At a given distance, the surface is
hotter than the interior by a factor of $\sim$3.
Though the interior in region ``II'' (see \S\ref{geomfan} above)
is not radially isothermal
as was found in the cruder analysis of CG97,
$T_i (a)$ does flatten slightly at these distances, as expected.
Deviations from a single power-law behavior for $T_{ds} (a)$ arise
from structure in $\varepsilon (\lambda)$.
For example, $T_{ds}$ declines slightly more steeply with $a$
for $0.5 \lesssim a_{\AU} \lesssim 2$ because
grains at these distances cool relatively efficiently
through silicate resonances at $10$--$20\mum$.

\subsection{Refined Standard SED}
\label{refstand}

\begin{figure}
\plotone{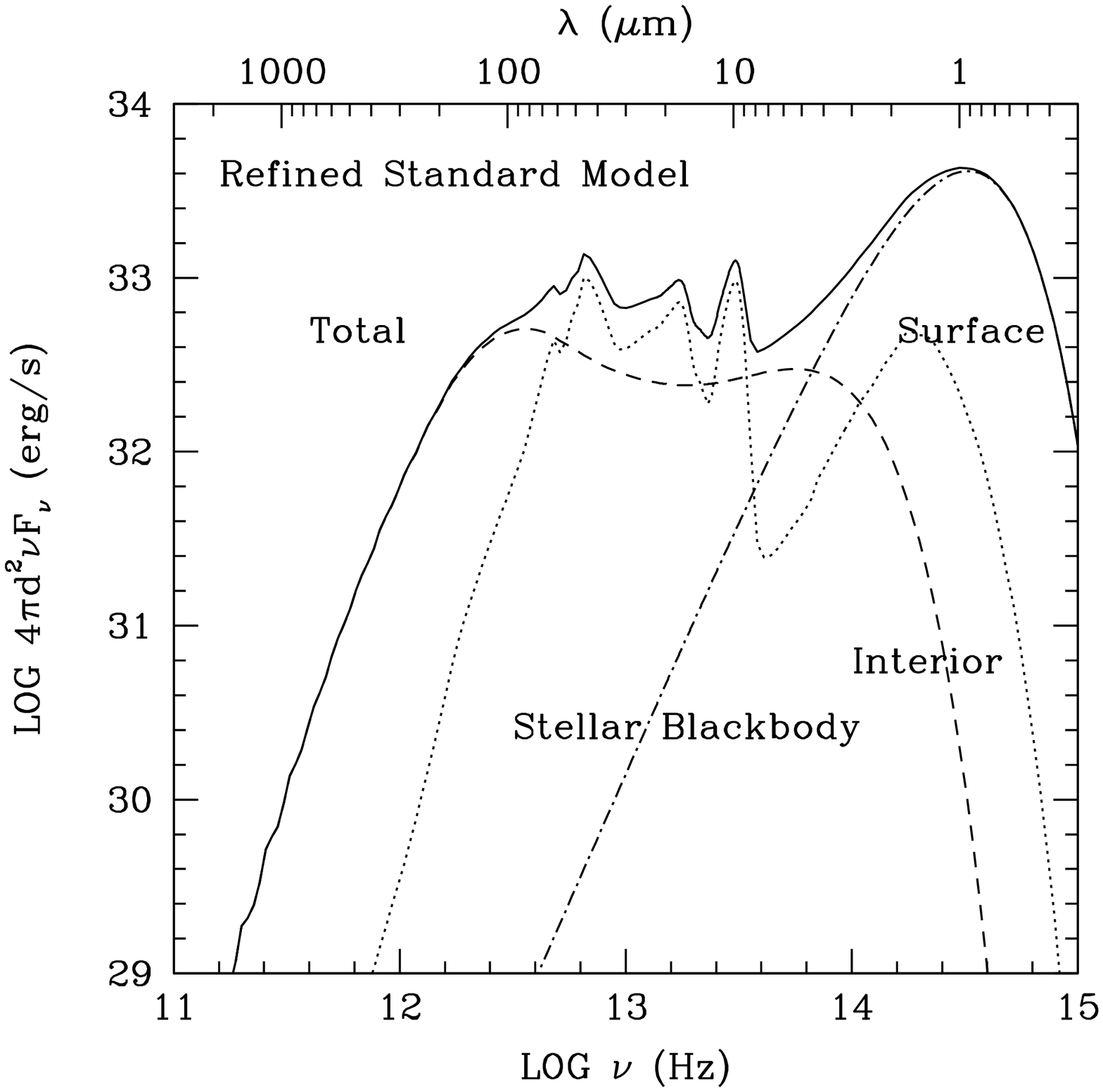}
\caption{Spectral energy distribution for our
refined standard model.
Resonant bands of superheated dust grains
in disk surface layers appear in emission. These include vibrational modes
in silicates at $10$ and $18\mum$, and lattice translational
modes in water ice at $45$ and $62\mum$. Absent are emission lines from
vibrational resonances in ice shortward of $10\mum$;
ice is not present in the disk surface inside
6 AU.$^{11}$ Note the dearth of emission from the surface between
$2$ and $8\mum$; amorphous olivine (MgFeSiO$_4$) grains
having $r \lesssim 1\mum$ are relatively transparent
at these wavelengths (see Figure \protect{\ref{qabs}}).
\label{sedfancy}}
\end{figure}

\placefigure{sedfancy}

The SED for our refined standard model is displayed
in Figure \ref{sedfancy}. The result shown here and that
of CG97 (see their Figure 6) differ primarily in the
emission from the superheated surface.
Here we have accounted in detail for the optical
properties of a few likely circumstellar grain materials.
Solid-state resonances of superheated
dust grains residing in disk surface layers
appear in emission. These include
vibrational modes in silicates at $10$ and $18\mum$,
and lattice translational modes in water ice at
$45$ and $62\mum$. Emission lines from
vibrational resonances in ice shortward of $10\mum$
are absent; ice is not present in the disk surface inside
6 AU.\footnote{It is conceivable that vibrational resonances
in water shortward of $10\mum$ may still appear in disk
spectra if silicate particles are hydrated. Such
signatures have been observed in spectra of CI chondritic meteorites (McSween,
Sears, \& Dodd 1988).} Note the dearth of emission from
the surface between $2$ and $8\mum$; amorphous olivine (MgFeSiO$_4$) grains
having $r \lesssim 1\mum$ are relatively transparent
at these wavelengths (see Figure \ref{qabs}).
The signature of this ``silicate transparent region,'' however,
is largely masked by emission from the optically thick
interior at small radius.
Furthermore, our computational experiments
demonstrate that including other types of particles
in the surface layer such as troilite (FeS)
serves to further fill in this transparent region.
The broad peak in disk surface
emission near $\lambda \sim 1.5 \mum$
arises from our pure iron particles
and iron impurities in our olivine particles.

\subsection{Dependence of SED on Input Parameters}
\label{paramexp}

\begin{figure}
\plotone{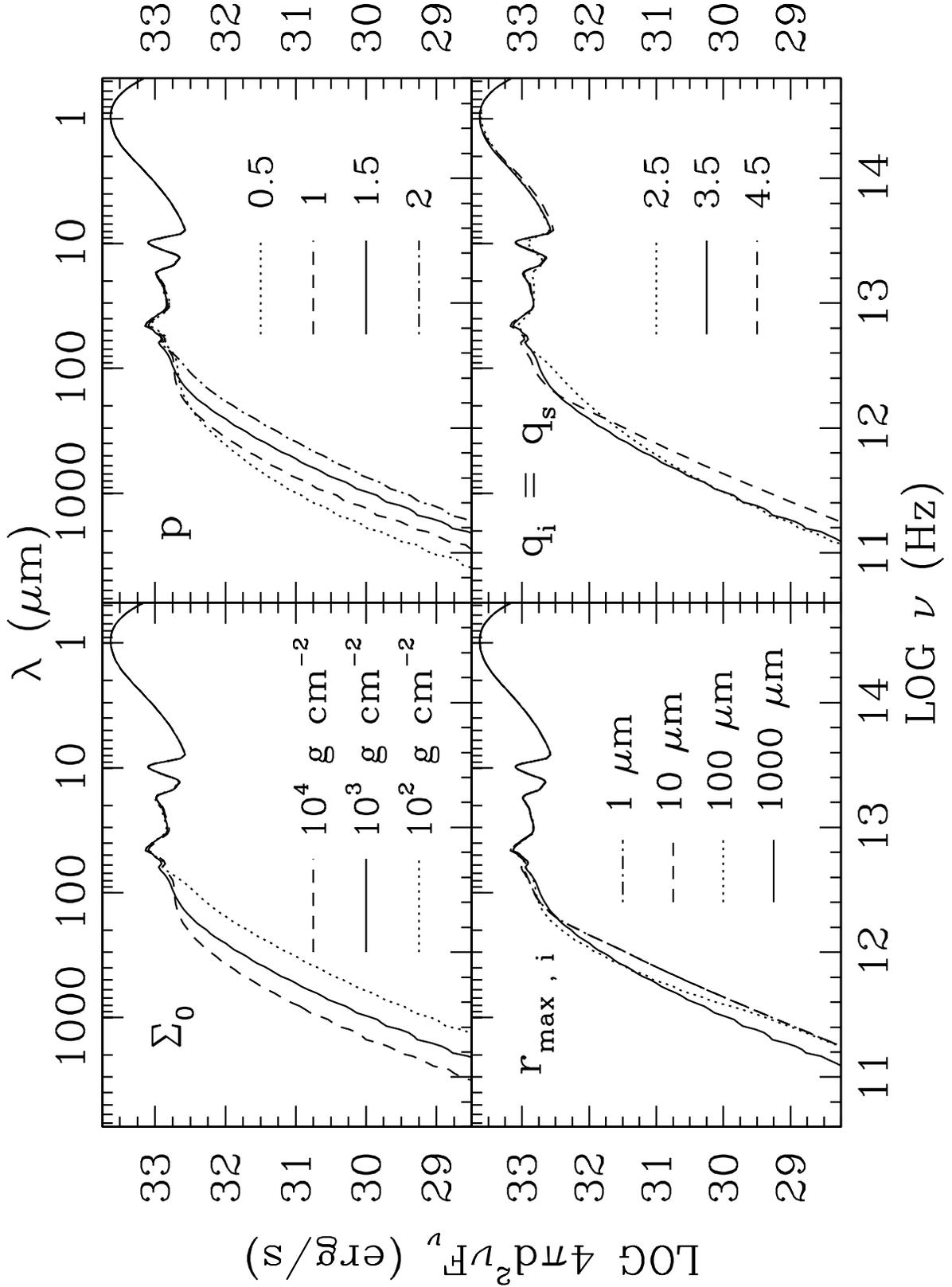}
\caption{Dependence of SED on input parameters
$\Sigma_0$, $p$, $r_{max,i}$, and $q_i=q_s$. In each panel,
the indicated parameter takes on several values while all
other parameters are fixed at their standard model values.
Grain size indices $q_i$ and $q_s$ are varied simultaneously
for compactness of presentation.
\label{parexp1}}
\end{figure}

\placefigure{parexp1}

\begin{figure}
\plotone{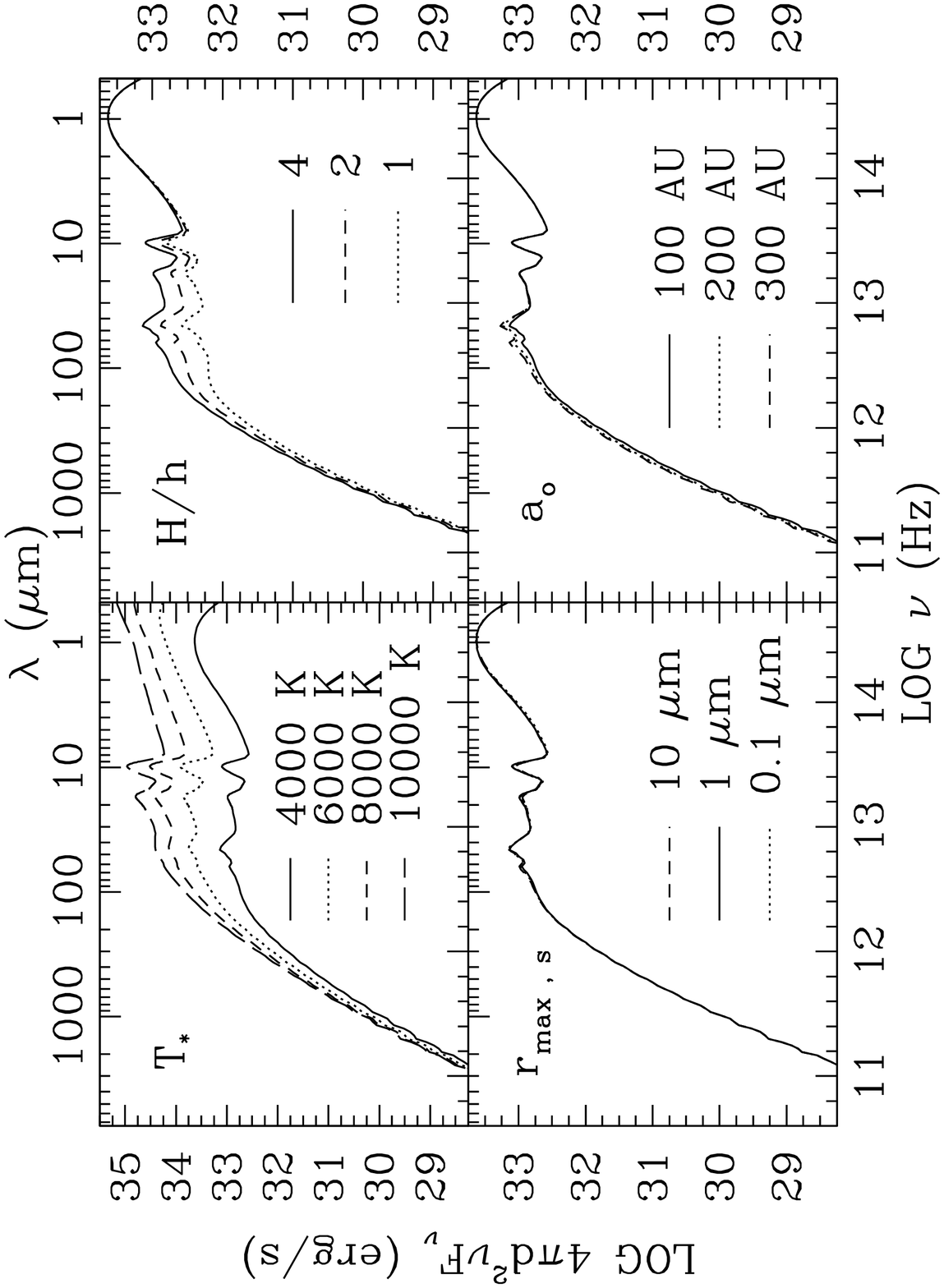}
\caption{Dependence of SED on input parameters
$T_*$, $H/h$, $r_{max,s}$, and $a_o$. In each panel, the
indicated parameter takes on several values while all other
parameters are fixed at their standard model values.
\label{parexp2}}
\end{figure}

\placefigure{parexp2}

In Figures \ref{parexp1} and \ref{parexp2}, we explore
the dependence of the SED on our input parameters.
In each panel, we vary the indicated parameter(s)
while fixing all other parameters at their standard
model values. All the variations are easily understood.
We observe the following behavior:

\begin{enumerate}

\item Millimeter-wave fluxes are most sensitive
to $\Sigma_0$, $p \equiv -d\ln \Sigma / d\ln a$,
$r_{max,i}$, and $q_i \equiv -d\ln N / d\ln r$.
The former two variables determine the amount
of mass in the cool disk interior at large radius;
the latter two variables affect the millimeter-wave opacity
in the disk interior.

\item Millimeter-wave SEDs for $r_{max,i} = 1$ and $10\mum$
are identical; these two cases belong to the
Rayleigh limit, in which absorptive cross-sections
are proportional to grain volume (Bohren \& Huffman 1983).
In this limit, millimeter-wave opacities are independent
of how the total condensable mass in water and silicates
is distributed with particle size.

\item As $r_{max,i}$ increases from 10 to $100\mum$,
the opacity at $\lambda = 100$--$600\mum$ in the
disk interior also increases, thereby enhancing
emission from the disk interior at those wavelengths.
A similar effect is seen as $r_{max,i}$ increases
from 100 to $1000\mum$.

\item For all values of $r_{max,i}$ considered, the spectral index
of the SED at $\lambda = 2$--$4$ mm equals
$n_{2-4} \equiv d\ln (\nu F_{\nu}) / d\ln \nu \equiv 3 + \beta_{\rm{eff}} =
4.6$.
The value of $\beta_{\rm{eff}} = 1.6$ equals the value
of $\beta$ for our ice-silicate grains, indicating
that radiation at these wavelengths emerges
from optically thin material.

\item For $q_s = q_i = 2.5$, most of the geometric cross-section
in our dust size distribution is concentrated in the largest
grains ($r_{max,s} \approx 1 \mum$
in the surface and $r_{max,i} \approx 1000 \mum$ in the interior).
Compared to the standard model, silicate emission
features from the surface at mid-infrared wavelengths
are weaker because the Rayleigh limit does not apply
for these large grains.
At millimeter wavelengths, $n_{2-4} = 4.3 \Rightarrow \beta_{\rm{eff}} = 1.3 <
\beta_{ice-sil} = 1.6$;
a substantial fraction of the mm-wavelength
emission arises from the disk interior made optically thick
by the increased number of mm-sized particles.
Note also the relative dearth of emission from $\lambda \approx 80$--$400\mum$
compared to the standard model, caused by fewer numbers of
$r \approx 15$--$70\mum$ sized particles in the disk interior and a
concomitant loss of interior opacity at these wavelengths.

\item For $q_s = q_i = 4.5$, most of the condensable mass
is concentrated in the smallest grains ($r_{min} \approx 0.01 \mum$).
Millimeter-wave opacities and therefore fluxes are
lower than for our standard model. Emission
at $\lambda \lesssim 100\mum$ does not differ
from standard model results because the Rayleigh limit
still applies for the most luminous grains in the surface;
shapes of the emissivity curves for silicate and ice-silicate
grains are independent of $r$ in the Rayleigh limit
(see Figure \ref{qabs}).

\item The radial locations of condensation
boundaries in disk surface layers move
outward approximately as $T_*^{(4+\beta)/2} \approx T_*^3$.
Consequently, as $T_*$ increases, surface emission
from water ice diminishes noticeably.

\item Reducing the height of the disk
photosphere by reducing the scaling parameter
$H/h$ (thereby crudely modelling the effects
of vertical settling of dust) lowers
the amount of stellar radiation intercepted
and reprocessed by the disk. Emission
at $\lambda \lesssim 200 \mum$
scales nearly linearly with $H/h$. Radiation
at these wavelengths arises from the optically thick interior (region ``I'')
and from the optically thin surface; for both regimes, $\nu F_{\nu}$
scales as $\sin \alpha$ which scales approximately as $H/h$.
Radiation at $\lambda \gtrsim 200\mum$---the Rayleigh-Jeans regime---is
less sensitive to $H/h$; here the approximate
scaling relation reads $\nu F_{\nu} \propto T_i \propto (\sin \alpha)^{0.25}
\propto (H/h)^{0.25}$.

\item Surface SEDs vary negligibly with
$r_{max,s}$. For our standard
slope of the size distribution ($q_s = 3.5$), silicate
and ice-silicate particles
having radii $r_{*} \approx 0.5 \mum$ absorb
the bulk of the radiation from the T Tauri star;
surface grains having $r \approx 0.2$--$1\mum$
are responsible for absorbing $\sim$50\% of
the incident stellar radiation.
If $r_{max,s} > r_*$, the SED is unchanged
because those grains having $r \gg r_*$ are
insufficiently numerous to be significant absorbers of radiation.
If $r_{max,s} < r_*$, the surface SED remains
unaltered because the Rayleigh limit still obtains.

\end{enumerate}

\section{Fitting SEDs of T Tauri and HAe Stars}
\label{fitherbig}

Our sample comprises 3 HAe and 2 T Tauri stars that
are (1) not known to harbor stellar companions, and (2)
not known to drive jets or to be surrounded by massive,
$\gtrsim 500\AU$-scale nebulosities that are better
described by spherical envelopes rather than by
flattened disks. In order of decreasing stellar
effective temperature, the sample stars
are MWC 480 (= HD 31648), HD 36112 (= MWC 758), CQ Tau (= HD 36910),
LkCa 15, and AA Tau.
For 4 of our sources (MWC 480, HD 36112, CQ Tau, and AA Tau),
medium-resolution ($\Delta \lambda = 0.2 \mum$) ISO LWS spectra
between 43 and $195\mum$ are available.
More detailed descriptions of the ISO data, including
how they were reduced and what gas phase spectroscopic
information they contain, are presented
elsewhere (Creech-Eakman, Chiang, van Dishoeck, \& Blake 2000).
All of our sources were too weak to be observed by the ISO Short Wavelength
Spectrometer.

The code that computes our refined standard model
is restricted to calculating SEDs for face-on disks.
A detailed study of how the SED varies with inclination, $i$,
has been given previously (CG99).
As described there, non-zero inclinations affect the
infrared SED in 2 principal ways: at moderate $i$,
by introducing a cosine $i$ variation in emission
arising from the optically thick interior, and
at extreme $i$, by blocking radiation
from the disk at small radius via the intervening
flared disk at large radius. The second effect
on the infrared SED is negligible for our sample stars.
Visual extinctions range from $A_V = 0.3$ mag (MWC 480)
to 2 mag (CQ Tau); these modest values imply that
inner disk regions are not significantly occulted
by flared outer disk edges at mid-to-far infrared wavelengths.
We estimate that the first effect introduces at most
a factor of 2.5 overestimation in our computed fluxes
between 2 and $8\mum$ where the SED is dominated
by emission from the optically thick interior.
For example, the disk inclination for CQ Tau has been
independently estimated to be $\sim$66$^{\circ}$, based
on photometric and polarimetric
variability at visible wavelengths (Natta \& Whitney 2000).
Such ``UXOR-type'' phenomena has been interpreted to arise
from clumps of dust in the flared disk surface sporadically obscuring
our line-of-sight to the central star.

For each source, a model SED is fitted to the ISO LWS scan (if
available), millimeter wavelength fluxes, and $\sim$3--25 $\mu{\rm m}$
photometric data. In 3 of the sources (AA Tau, CQ Tau, and MWC 480),
ISO fluxes are greater than corresponding IRAS (Infrared Astronomical
Satellite) fluxes at 60 and $100\mum$ by factors of $\sim$2--3.
The origin of the discrepancies is not known.
Where there are discrepancies, preference is given to the
ISO LWS data for which the beam area is $\sim$2 times smaller
than that of IRAS. For HD 36112, there is excellent
agreement between ISO and IRAS.
Preference is given also to the central
portions of the ISO scans between 50 and $170 \mum$ where individual detectors
overlap in wavelength coverage and measured fluxes are consequently
more reliable.

\begin{figure}
\plotone{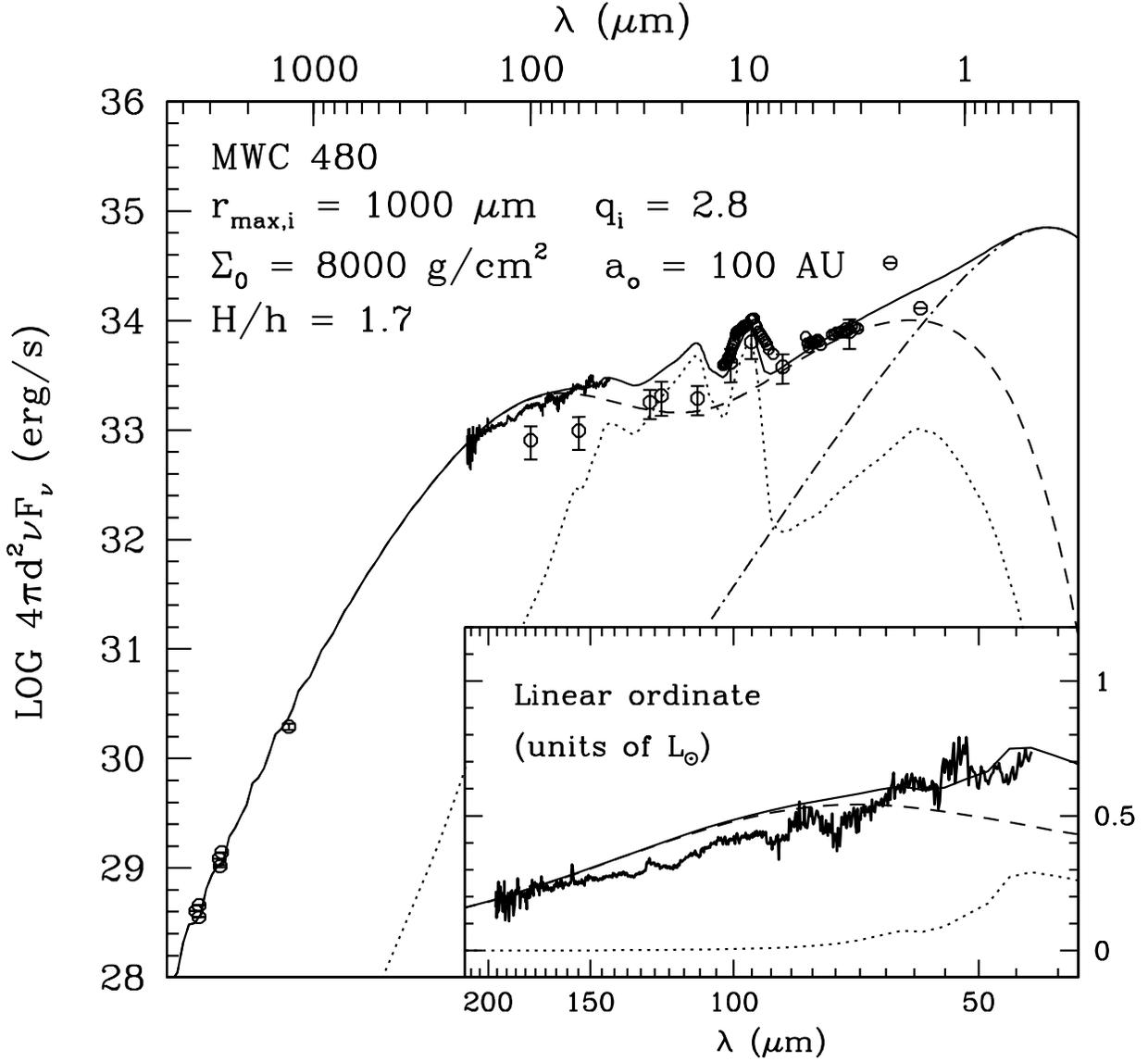}
\caption{Refined 2-layer model fitted
to data for MWC 480. Dotted lines denote the contribution
from surface layers. Dashed lines denote the
contribution from the disk interior. The inset plot
with linear ordinate magnifies the spectral region spanned by ISO LWS.
Photometric data are taken from Mannings \& Sargent (1997),
the color-corrected IRAS Point Source Catalog (1988),
Sitko et al.~(1999), Thi et al.~(2000), and Qi (2000).
\label{mwc}}
\end{figure}

\placefigure{mwc}

\begin{figure}
\plotone{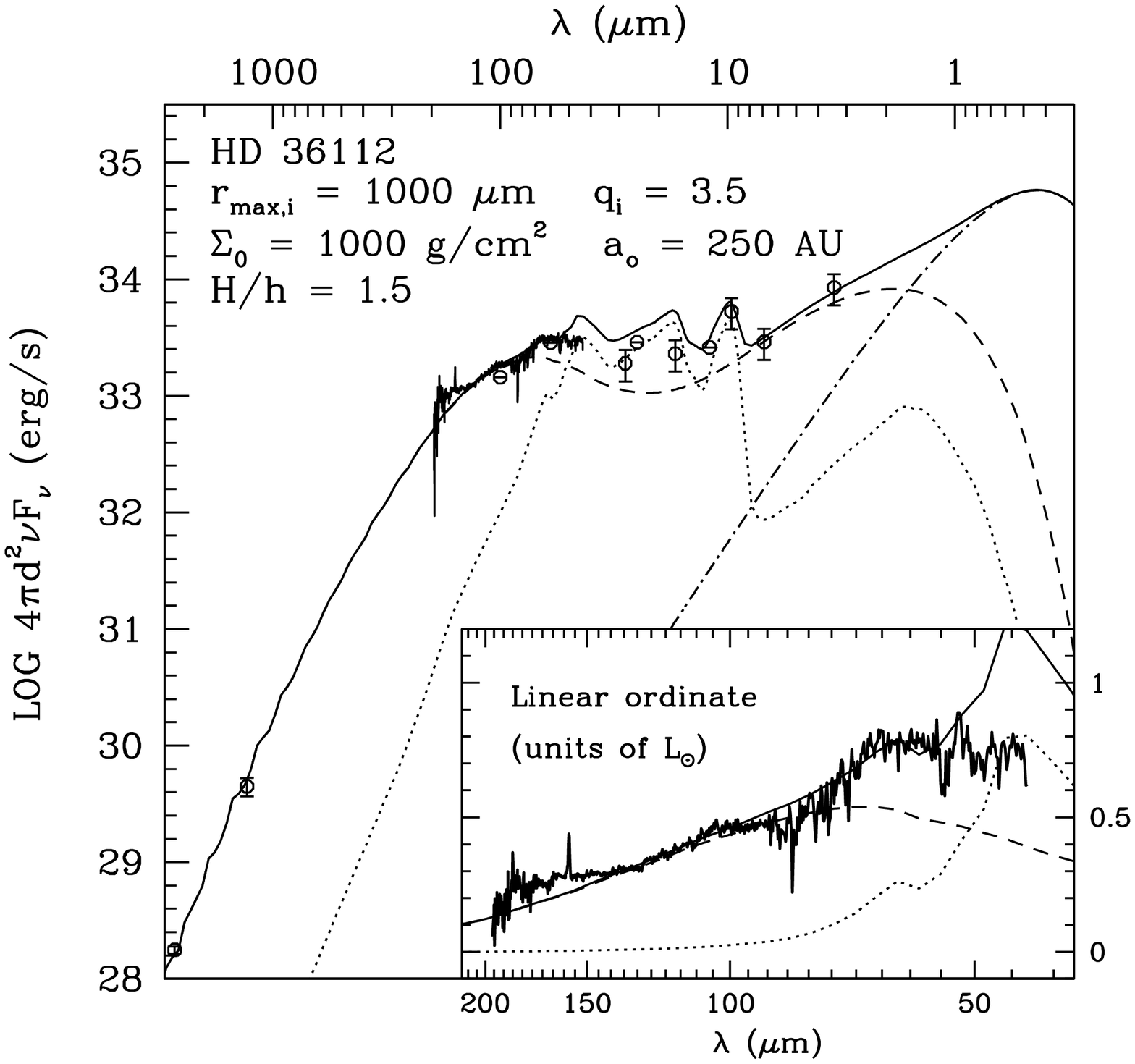}
\caption{Refined 2-layer model fitted
to data for HD 36112. Photometric data
are taken from Mannings \& Sargent (1997),
the color-corrected IRAS Point Source Catalog (1988),
and Thi et al.~(2000).
\label{hdsed}}
\end{figure}

\placefigure{hdsed}

\begin{figure}
\plotone{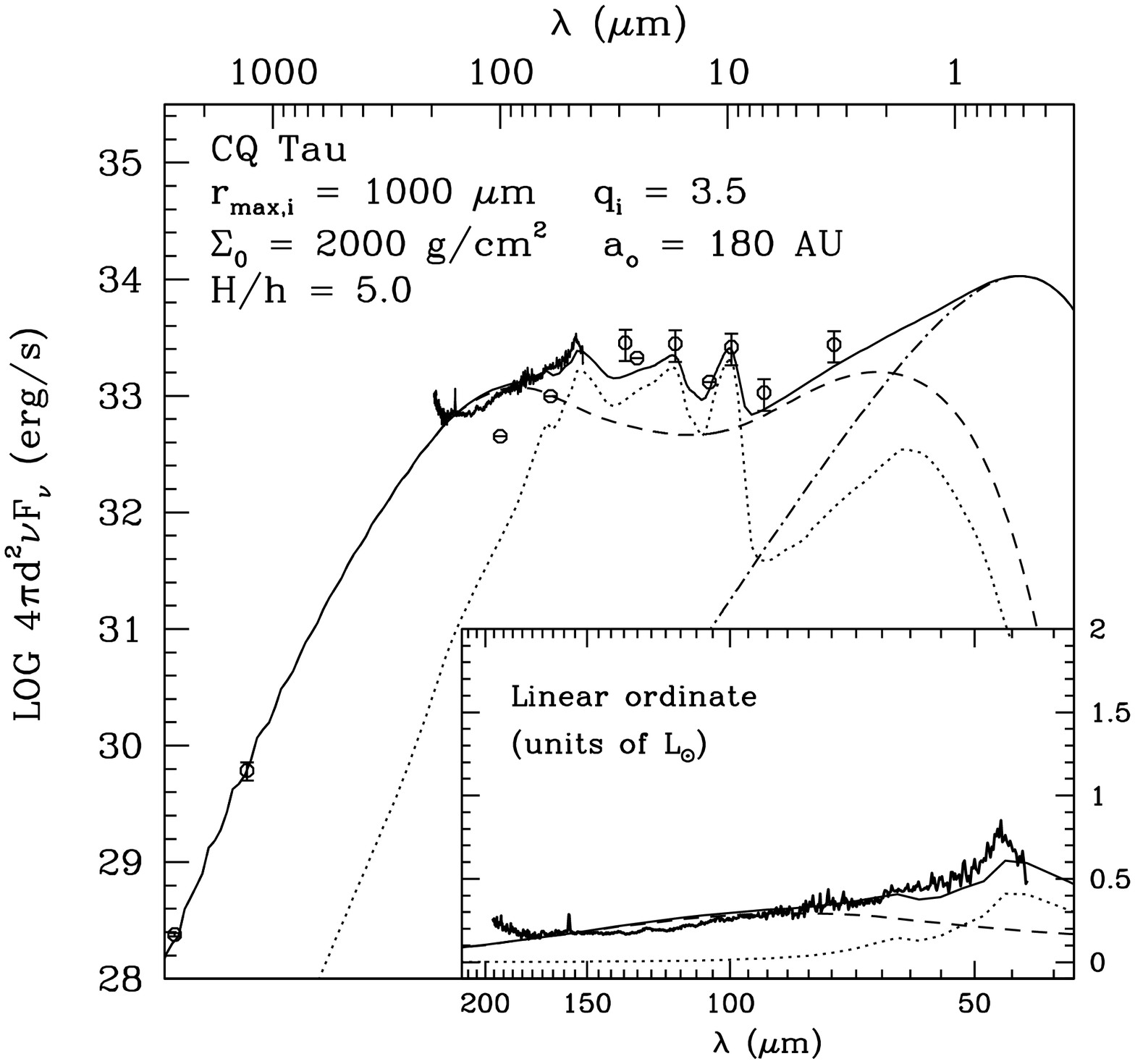}
\caption{Refined 2-layer model fitted to
data for CQ Tau. Photometric data
are taken from Mannings \& Sargent (1997),
the color-corrected IRAS Point Source Catalog (1988),
and Thi et al.~(2000).
\label{cqsed}}
\end{figure}

\placefigure{cqsed}

\begin{figure}
\plotone{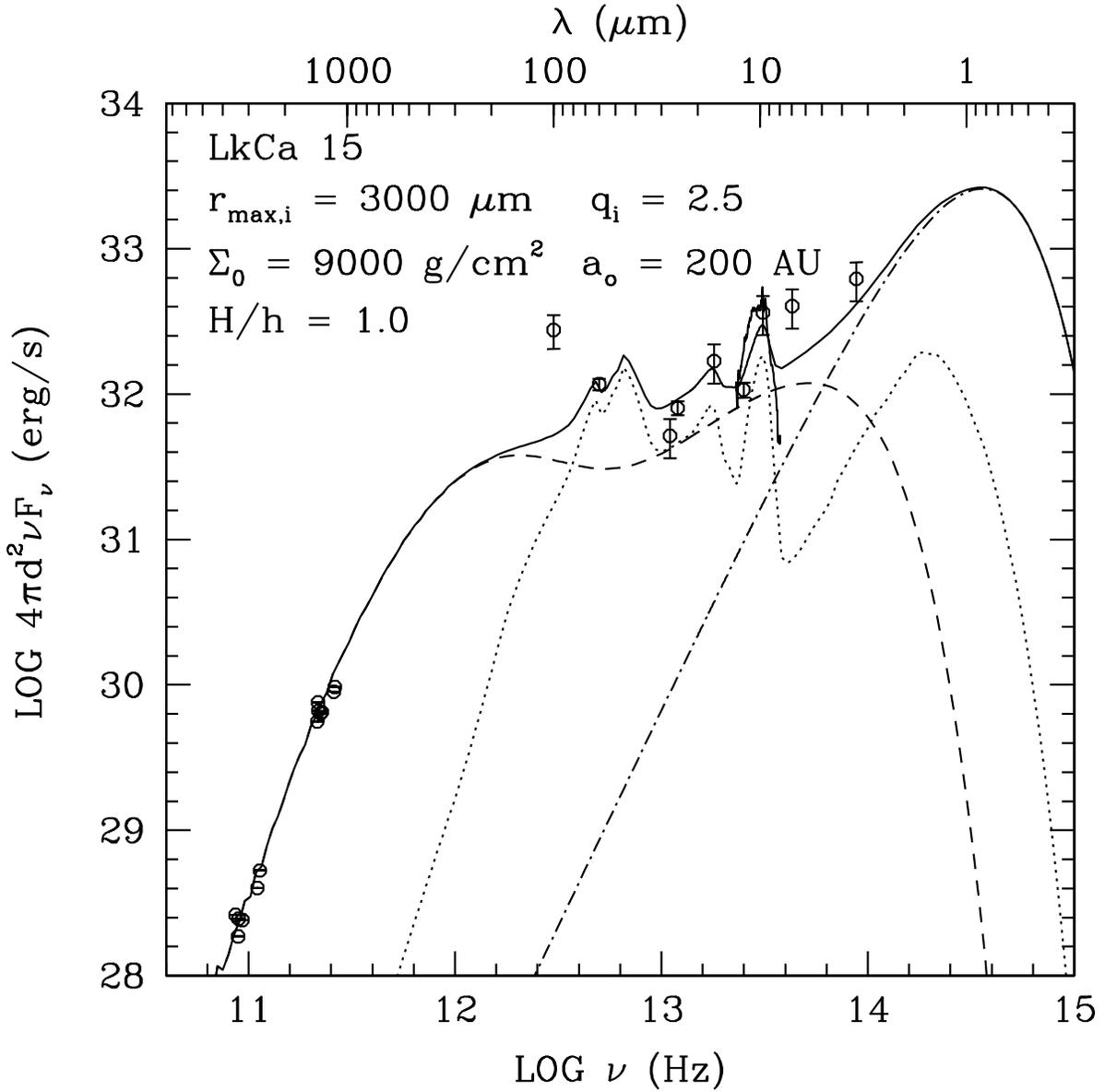}
\caption{Refined 2-layer model fitted to
data for LkCa 15. Photometric data are
taken from the color-corrected IRAS Point Source
Catalog (1988), Thi et al. (2000), and Qi (2000).
\label{lkca}}
\end{figure}

\placefigure{lkca}

\begin{figure}
\plotone{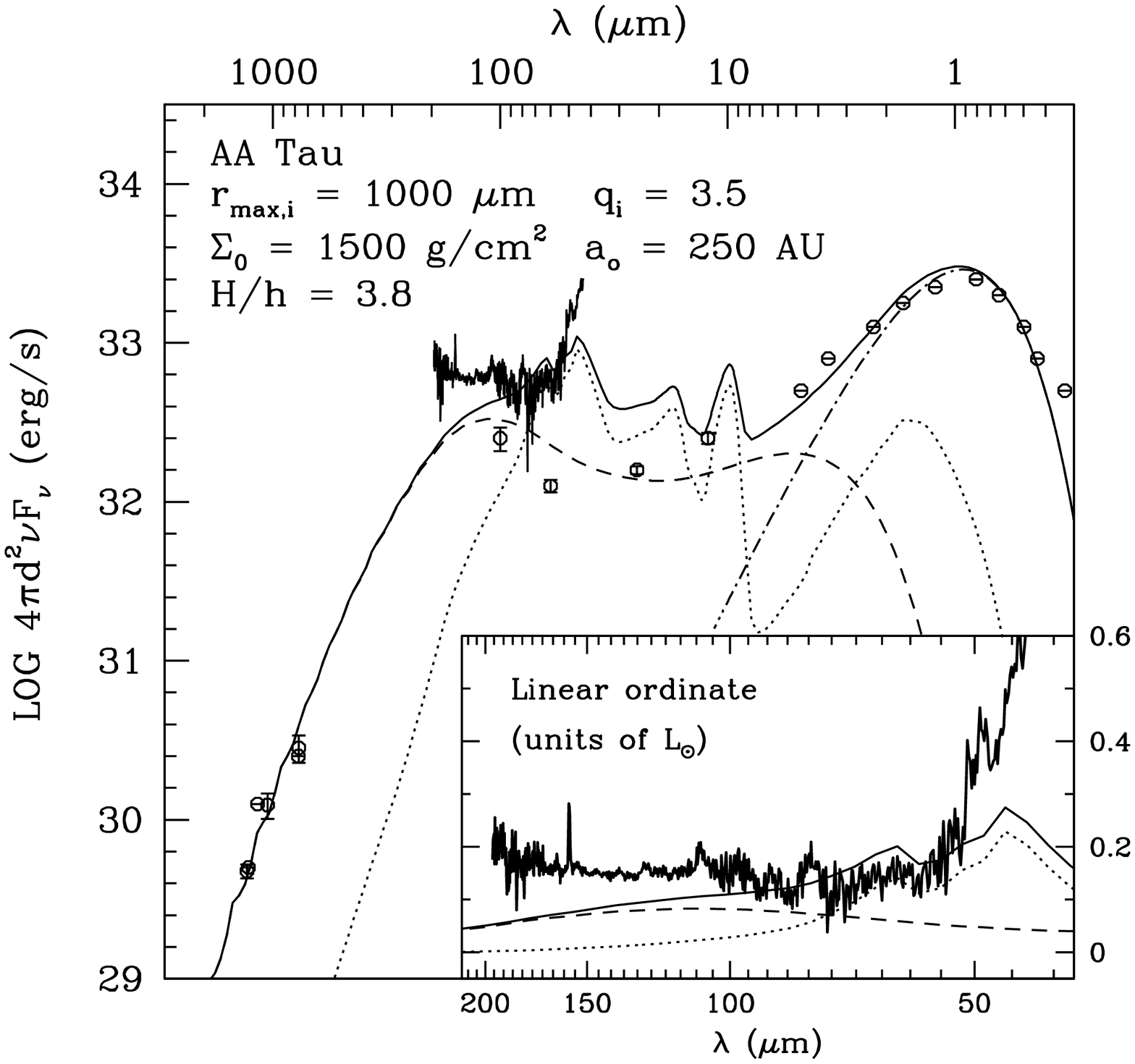}
\caption{Refined 2-layer model fitted to
data for AA Tau. Photometric data are taken
from Beckwith et al.~(1990), Beckwith \& Sargent (1991),
and Dutrey et al.~(1996).
\label{aa}}
\end{figure}

\placefigure{aa}

In fitting the SEDs, we fix $q_s = 3.5$.
Smaller values ($q_s < 3$) seem unlikely
since they would imply that the largest grains,
which tend to settle out of surface layers most quickly,
dominate the geometric cross-section. The SED is
insensitive to larger values ($q_s > 4$), as shown
in \S\ref{paramexp}. We also
fix $r_{max,s} = 1 \mum$ and $p = 1.5$ for all models.
Section \ref{paramexp} demonstrates
that the SED is insensitive to $r_{max,s}$ once
$q_s$ is fixed at 3.5, and that $\Sigma_0$ and $p$
affect the SED in similar ways.

\begin{deluxetable}{cccccc}
\tablewidth{6in}
\tablecaption{Fitted Parameters\tablenotemark{a}~ of Herbig Ae and T Tauri
Star/Disk Systems\tablenotemark{b}\label{fitpar1}}
\tablehead{
\colhead{Parameter}      & \colhead{MWC 480} & \colhead{HD 36112} & \colhead{CQ
Tau} & \colhead{LkCa 15} & \colhead{AA Tau}}
\startdata
$T_*$ (K) & 8890 & 8465 & 7130 & 4395 & 4000 \\
$R_* (R_{\odot})$ & 2.1 & 2.1 & 1.27 & 1.64 & 2.1 \\
$M_* (M_{\odot})$ & 2.3 & 2.2 & 1.7 & 1.0 & 0.67 \\
$d$ (pc) & 140 & 150 & 100 & 140 & 140 \\
$\Sigma_0 (\rm{g}\, \cm^{-2})^{\rm c}$ & 8000 & 1000 & 2000 & 9000 & 1500 \\
$a_o$ (AU)$^{\rm c}$ & 100 & 250 & 180 & 200 & 250 \\
$H/h$ & 1.7 & 1.5 & 5.0 & 1.0 & 3.8 \\
$q_i$$^{\rm c}$ & 2.8 & 3.5 & 3.5 & 2.5 & 3.5 \\
$r_{max,i}$ $(\mu\rm{m})\tablenotemark{c}$ & 1000 & 1000 & 1000 & 3000 & 1000
\\
$M_{\rm{DISK}} (M_{\odot})$$^{\rm d}$ & 0.11 & 0.02 & 0.04 & 0.18 & 0.03 \\
$H(a_o) / a_o$ $\tablenotemark{d}$ & 0.13 & 0.16 & 0.45\tablenotemark{e} & 0.09
& 0.58 \\
\enddata
\tablenotetext{a}{{\footnotesize For all sources, we fix $q_s = 3.5$,
$r_{max,s} = 1\mum$,
and $p = 1.5$. See text for rationale.}}
\tablenotetext{b}{{\footnotesize Stellar parameters and distances for HAe stars
are
taken from Mannings \& Sargent (1997), except for CQ Tau
for which $R_*$ and $d$ are normalized to the Hipparcos distance.
Stellar parameters
and distances for T Tauri stars are taken from
Beckwith et al.~(1991) and Webb et al.~(1999).}}
\tablenotetext{c}{{\footnotesize The continuum SED is largely degenerate with
respect to
simultaneous changes in $\Sigma_0$, $r_{max,i}$, $q_i$, and $a_o$. The values
shown here are not uniquely constrained.}}
\tablenotetext{d}{{\footnotesize Total masses (in gas and dust) and maximum
aspect ratios of fitted disks are derived quantities and not
input parameters. As discussed in \S\ref{degenesec}, the
total disk mass depends sensitively on the {\it a priori} unknown
millimeter-wave opacity and could vary by an order of magnitude.}}
\tablenotetext{e}{{\footnotesize Our fitted maximum aspect ratio for CQ Tau and
the
modest visual extinction to this star (2 mag) imply a disk
inclination $i \lesssim 66^{\circ}$ ($i = 0$ for face-on views).
This is consistent with the independent estimation of $i \approx 66^{\circ}$ by
Natta \& Whitney (2000) based on UXOR-type variability at
visible wavelengths.}}
\end{deluxetable}

Observations and fitted theoretical models
are displayed in Figures \ref{mwc}--\ref{aa}.
Table \ref{fitpar1} contains the fitted parameters for
our sample. The fits are intended to be illustrative; no attempt
is made to minimize fit deviations in a formal,
statistical sense. The results of such an analysis would not be
very meaningful anyway, since the SED tends to be degenerate
with respect to simultaneous changes in several~of~the~parameters,
as we discuss in \S\ref{degenesec}. In any case, our
fitted outer disk truncation radii
for CQ Tau, HD 36112 (MWC 758), and MWC 480 are in accord
with upper limits based on $\lambda = 2.7 \mm$ continuum images
taken by Mannings \& Sargent (1997).

With few exceptions, the
agreement between models and observations is good to within
a factor of 2, and serves as further evidence that simple
reprocessing of central starlight by flared disks adequately explains
the infrared excesses of these systems. We are aware that this is
not the conclusion of Miroshnichenko et al.~(1999, hereafter MIVE99), who
require the presence of tenuous envelopes having radii
$\sim$$1000\AU$ to heat embedded HAe disks to temperatures greater
than those predicted by the classical $T \propto a^{-3/4}$ law.
This idea was first elucidated by Natta (1993)
in the context of flat spectrum T Tauri stars. While some HAe systems
do exhibit large scale nebulosities whose sizes as functions
of wavelength vary in accord with the calculations of MIVE99,
we disagree with the statement that disks are incapable of explaining
the SEDs of HAe systems without heating from an envelope.
Non-classical temperature laws follow naturally from
passive reprocessing of starlight
by hydrostatically flared disks (e.g., Kenyon \& Hartmann 1987;
CG97; this paper).

Moreover, MIVE99 claim that radiation from disk surface layers
as envisioned by CG97 do not contribute substantially to SEDs
of HAe stars MWC 480 and CQ Tau. However, in Figures \ref{mwc}
and \ref{cqsed}, our model fits to these 2 sources indicate that
(1) emission from optically thin disk surface layers dominates
emission from the optically thick disk interior between 10 and $\sim$$50\mum$,
and (2) surface layer emission naturally explains the observed
infrared excesses, in particular the presence of silicate and
ice emission bands (see \S\ref{isel} for a more complete
discussion of these bands).

Certainly the SED alone does not furnish sufficient information to uniquely
constrain the geometry of dust surrounding HAe stars.
We agree with MIVE99 that the presence of envelopes is an
important possibility to consider for all HAe stars,
and that for sources such as AB Aur their existence
is persuasively implicated by imaging data at multiple
wavelengths (see \S2 of MIVE99). We simply wish to emphasize
here that the SED alone does not rule out isolated disks
heated by their central stars, and that envelopes
are not the only way to achieve extra heating of the disk.


\subsection{Degeneracy between Disk Mass and Grain Size Distribution}
\label{degenesec}

\begin{figure}
\plotone{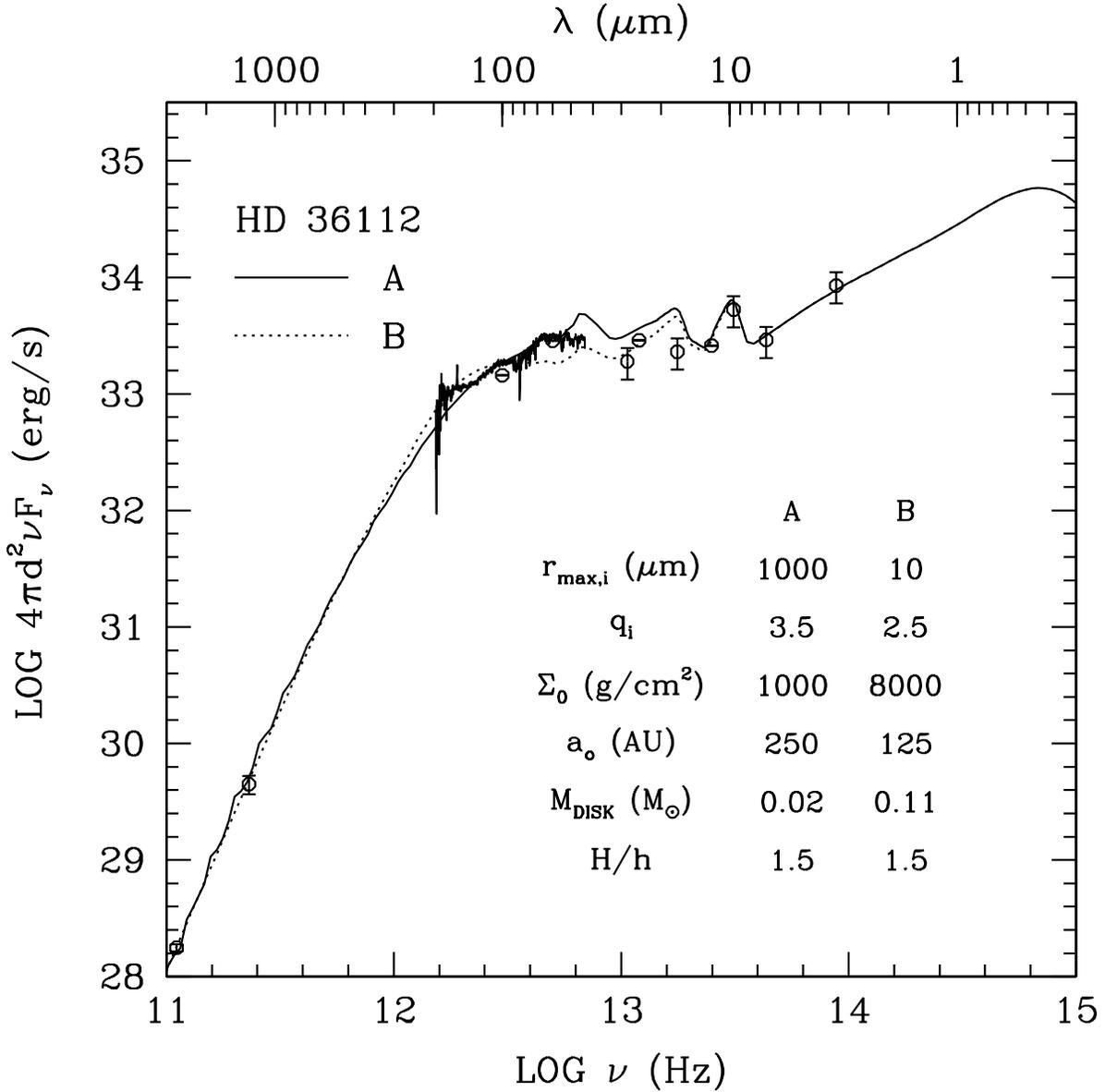}
\caption{Degeneracy between surface density
and grain size for HD 36112.
Two models having different sets of parameters
are fitted to the same dataset. The disk mass, $M_{\rm{DISK}}$,
is not an explicitly inputted parameter but is derived
from $\Sigma_0$, $p = 1.5$, $a_o$, and $a_i = 2 R_*$.
Smaller grain sizes (i.e., smaller
millimeter-wave opacities) may be traded for larger disk masses
to achieve similar mm-wave optical depths and
fits of comparable quality. Note, however, that both
models require $H/h = 1.5$ to match the overall level of infrared
excess at $\lambda \lesssim 100\mum$.
\label{degen}}
\end{figure}

As might be gleaned from Figures \ref{parexp1} and \ref{parexp2},
the values for $\Sigma_0$, $p$, $r_{max,i}$, $q_i$, and $a_o$
presented in Table \ref{fitpar1} cannot be
uniquely constrained by the continuum SED alone.
We display one degenerate combination in Figure \ref{degen},
where 2 models using 2 different sets of
parameters are fitted to the observed data for HD 36112.
We feel that the fits are of comparable quality, given
the crudeness of our 2-layer model.
It follows that the total disk mass (in dust) is highly
uncertain; in fact, our two models for HD 36112 differ
in their total dust mass by a factor of $\sim$6.
One relies on the optical thinness of the disk
interior at millimeter wavelengths to
measure the disk mass in dust: $M_{\rm{DISK}} \propto F_{\rm{mm}} \propto
\tau_{\rm{mm}} \propto \Sigma \kappa_i$. The problem is that the SED
alone cannot disentangle $\Sigma$ from $\kappa_i$, which in
turn depends on $q_i$ and $r_{max,i}$.

\placefigure{degen}

The two models in Figure \ref{degen} yield
significantly different results for the mass
distribution in space, however. Upcoming spatially resolved observations
at millimeter wavelengths will help to break this
degeneracy between grain size (i.e., millimeter-wave opacity)
and disk surface density (Beckwith, Henning,
\& Nakagawa 2000).

\subsection{Evidence for Dust Settling}

The one disk parameter which appears to be most uniquely
constrained is $H/h$, the height of the disk photosphere in
units of the gas scale height. Its value is roughly proportional to
the overall level of infrared excess at $\lambda \lesssim 100\mum$.
CQ Tau exhibits $H/h = 5.0$, a value appropriate for gas
and dust that are well-mixed in interstellar proportions.
Many of our sources, however, are fitted with significantly
lower values between 1 and 4. We interpret these low values
to mean that dust in disk surface layers has settled vertically
towards the midplane. This was expected from the analysis of CG97
(see their \S3.3), and reinforces similar conclusions by
D'Alessio et al.~(1999).

The overall level of infrared excess at $\lambda \lesssim 100\mum$ can also
decrease with increasingly edge-on viewing angles,
as the central portions of the disk are increasingly
hidden from view by the flared outer ``wall'' (CG99).
Inclination effects appear insufficient
to explain the relatively low infrared excesses exhibited by
sources here, however, because their visual extinctions are
$\lesssim 1$ mag and their near-to-mid-IR fluxes are
not lower than their far-IR fluxes; the reverse would be true
for significantly inclined sources. Using the results
from CG99, we estimate that
infrared excesses may be depressed by a factor of $\sim$1.5
for our sources due to non-zero inclination; the
suppression factor due to lower values of $H/h$ is $\sim$2--4.

\subsection{Ice and Silicate Emission Lines}
\label{isel}

Observational evidence for silicate emission at $10\mum$
from the superheated surface exists for all
of our sources except AA Tau. In cases where medium-resolution
spectra exist (MWC 480, LkCa 15),
the ``trapezoidal'' shape of the observed emission feature
is imperfectly fitted by our model; this
indicates that actual surface layer silicates have
allotropic states (crystalline vs. amorphous)
and compositions (pyroxene vs. olivine, and
Fe:Mg ratios) slightly different from the amorphous
MgFeSiO$_4$ that we employ. Sitko et al.~(1999) reproduce
the ``trapezoidal'' shape of the $10\mum$ emission feature
using an admixture of
amorphous olivine, amorphous enstatite (a pyroxene),
and crystalline olivine, in roughly equal proportions.
The amplitudes of the observed
emission bands indicate that the dominantly absorbing (emitting)
silicates in the surface layer have sizes $r\lesssim 1\mum$, well
inside the Rayleigh limit.

Observational evidence for water ice emission
at $\sim$$45\mum$ is present in ISO scans of
2 of the coolest stars, CQ Tau and AA Tau.
The feature at $45\mum$ represents the translational mode in water ice having
the highest oscillator strength.
If we examine only the ISO data in the magnified
inset plots of Figures \ref{mwc}, \ref{hdsed}, \ref{cqsed}, and \ref{aa},
and ignore the model fits, there appears to be a
trend of increasing $\sim$$45\mum$ flux
relative to $\sim$$55\mum$ continuum flux
with decreasing stellar effective temperature.
This behavior accords with the trend noted
in \S\ref{paramexp} (see Figure \ref{parexp2}),
whereby the amount of ice present in disk surface layers decreases rapidly
with increasing $T_*$.

Notwithstanding this qualitative agreement,
the model fits to water ice emission bands
(or lack thereof) in the ISO spectra require substantial improvement.
The best fitted cases include (1) CQ Tau, for which there
is even observational evidence of an additional translational
band at $60$--$65\mum$ which our model reproduces,
and (2) MWC 480, for which the high stellar temperature
and the relatively small outer truncation radius of the disk
suppress water ice emission to levels in approximate
accord with observations.
However, the observed width of the $45\mum$ band
in CQ Tau is narrower than what our model predicts.
The discordancy of emission line shapes at $45\mum$
is yet stronger in the cases of AA Tau and HD 36112.
In the latter case, water ice emission
is predicted by the model but is not observed.
See, however, Figure \ref{degen}
for an alternative model in which we substantially
reduce the outer disk radius of HD 36112, thereby
reducing the $45\mum$ flux.

Accurate reproduction of observed solid-state emission features
from water is hampered by a number
of difficulties. Aside from the crudeness
of our 2-layer radiative transfer scheme,
these obstacles include (1) uncertainties in the photospheric abundances
of water relative to silicates (we have assumed
cosmic abundances with 50\% of the oxygen tied
up in water and 100\% of the iron locked in refractory grains);
(2) uncertainties in how ice is distributed with
particle size (we have assumed a constant
fractional radial thickness of the ice mantle relative
to the radius of the silicate core for a power-law
distribution of core radii);
(3) the probable presence of impurities in water ice that can shift
band positions and widths; and (4) incompleteness
of laboratory data for the optical constants of
a cosmic mixture of ices in various allotropic states
at wavelengths longward of $100\mum$. Improving the fits
by attacking these problems is beyond the scope of our
present, exploratory work.

Finally, we note the existence of several apparent
emission bands in the ISO spectra that we are unable
to identify. Most prominent among these is
a broad peak near $80\mum$ in scans of AA Tau,
MWC 480, and possibly CQ Tau and HD 36112.
No resonance at $80\mum$ exists for amorphous silicates
(J\"{a}ger et al.~1994); nor is such a resonance
measured for the crystalline silicates studied by J\"{a}ger et al.~(1998).
However, these latter authors also show that peak positions of a given
vibrational mode shift towards longer wavelengths with increasing
iron-to-magnesium content (higher effective vibrating masses).
We propose that the $80\mum$ feature
is caused by a translational mode in crystalline olivine having an
Fe:Mg ratio intermediate between that of ``natural olivine''
(Mg$_{1.96}$Fe$_{0.04}$SiO$_4$) and ``natural hortonolite''
(Mg$_{1.1}$Fe$_{0.9}$SiO$_4$)
(J\"{a}ger et al.~1998, see their Table 3).
If this is the case, we would expect an associated crystalline
silicate translational band to appear near $51\mum$; indeed, such an
emission line does appear in ISO scans of AA Tau, MWC 480, and HD 36112.
With regards to these and other perceived emission bands in
the ISO data, however, the possibility of instrumental
error must unfortunately be kept in mind (see \S4.2.2 of
Creech-Eakman et al.~2000).

\subsection{Near-Infrared Excesses}
Our model near-infrared fluxes shortward of $10\mum$ are occasionally
underestimated in MWC 480,
LkCa 15, and AA Tau. If we adopt the disk inclination
of $i \approx 66^{\circ}$ for CQ Tau (Natta \& Whitney 2000)
and the resultant lowering of its model near-IR fluxes by
a factor of 2.5 (see \S\ref{fitherbig}), then this star's observed near-IR
fluxes are also inadequately explained. The relative dearth of
surface layer emission arises partly
because silicate particles are particularly
transparent in this wavelength regime that
exists between the vibrational resonances
near $10\mum$ and absorption due to iron
impurities near $1\mum$. It seems possible
that thermally-spiked emission from polycyclic
aromatic hydrocarbons (PAHs) in disk surface layers
may help to fill in this transparent region
of the SED. The strongest resonances
are due to C-H and C-C stretching and
bending modes at 3.3, 6.2, 7.7, 8.6 and
$11.3\mum$ (Draine 1995). Upcoming high spectral
resolution observations with SOFIA
(Stratospheric Observatory For Infrared Astronomy)
can test this hypothesis. We note, however, that
Natta, Prusti, \& Krugel (1993) and
Bouwman et al.~(2000) argue against this explanation
for HAe stars AB Aur and HD 163296 based
on the need for unrealistically high abundances of PAHs
and on available ISO Short Wavelength Spectrometer data.

Other neglected but possibly relevant contributions
to near-IR excesses include (1) active accretion, which is likely
to play an important role at disk radii
inside a few AU, (2) reflected starlight off the disk
surface, and (3) the possibility of dust components in
addition to the disk, e.g., an optically thin, spherically
distributed cloud of silicate/iron grains within a few
AU of the star (MIVE99; Bouwman et al. 2000).

\section{Summary}
\label{summ}

In this work, we have constructed improved versions
of 2-layer passive disk models by Chiang \& Goldreich (1997).
These improvements include an explicit accounting of
grain size distributions and grain compositions,
and numerical solution of the equations
of radiative and hydrostatic equilibrium under
the original 2-layer approximation. We have
explored how the SED varies in input parameter
space and applied our models to observations
of 5 T Tauri and Herbig Ae stars. Our principal
conclusions are as follows:

\begin{enumerate}

\item Hydrostatically flared, passive disks having
masses of $\sim$0.01--$0.1 M_{\odot}$ and radii of
$\sim$100--250 AU adequately explain the infrared-to-millimeter
wavelength excesses of our sample classical T Tauri and HAe stars.
Unambiguous determination of the geometry of circumstellar
dust requires, however, spatially resolved images.
Maps from near-infrared to millimeter wavelengths
generated by the Atacama Large Millimeter Array (ALMA),
the Space Infrared Telescope Facility (SIRTF), and the
Next Generation Space Telescope (NGST) will help to break
degeneracies inherent in SEDs between, e.g., those of
disks and envelopes.

\item Solid-state spectral features in the mid-infrared
($\lambda = 5$--$60\mum$) appear in emission from face-on disks.
These emission features arise from ``disk atmospheric grains'': grains
in disk surface layers that are directly irradiated by central
starlight. The strongest resonances include the $10\mum$ peak from
surface silicates at stellocentric distances of a few AU,
and the $45\mum$ peak from surface water ice at distances of $\sim$100 AU.
The strengths of these emission bands relative
to that of the adjacent continuum depend on (a) the sizes
of atmospheric grains that absorb the bulk
of the stellar radiation, and (b) the disk viewing
geometry. If atmospheric grain sizes are within
the Rayleigh limit ($2\pi r / \lambda \lesssim 1$),
emission band amplitudes saturate relative to the continuum.
As atmospheric grain sizes increase beyond
the Rayleigh limit ($2\pi r / \lambda \rightarrow \infty$),
emission band amplitudes decrease. In addition, as the
disk is viewed at increasingly edge-on inclinations,
emission bands tend to go into absorption (CG99).

\item Values for $\Sigma_0$, $p$, $r_{max,i}$, $q_i$, and $a_o$
influence the SED most at wavelengths longward of $100\mum$.
Their values for a given source, however, cannot be uniquely constrained
by the millimeter-wave SED alone. One relies
on the optical thinness of the disk interior at millimeter wavelengths to
measure the disk mass in dust: $M_{\rm{DISK}} \propto F_{\rm{mm}} \propto
\tau_{\rm{mm}} \propto \Sigma \kappa_i$. The problem is that the SED
alone cannot disentangle $\Sigma$ (which depends
on $\Sigma_0$, $p$, and $a_o$) from $\kappa_i$ (which in
turn depends on $q_i$ and $r_{max,i}$).
Spatially resolved millimeter-wave maps can help to
break the degeneracy between interior grain size and disk mass.

\item The one disk parameter which appears to be most uniquely
constrained by the SED is $H/h$, the height of the disk photosphere in
units of the gas scale height. Its value is roughly proportional to
the overall level of infrared excess at $\lambda \lesssim 100\mum$.
CQ Tau exhibits $H/h = 5.0$, a value appropriate for gas
and dust that are well-mixed in interstellar proportions.
Our other sources---MWC 480, HD 36112, LkCa 15, and AA Tau---are
fitted with significantly lower values between 1 and 4. We
interpret these low values to mean that atmospheric grains
in disk surface layers have settled vertically
towards the midplane. For standard disk parameters,
the time required for a 0.1 micron-sized grain to
settle from $H = 4h$ to $H = 0$ is $8 \times 10^6 \yr$
in the absence of vertical gas flow; from $H = 4h$ to $H = h$,
the required time is $8 \times 10^5 \yr$ (CG97; Creech-Eakman et al. 2000).
Both these times are of the same order of magnitude
as the estimated stellar ages. The actual amount
of photospheric settling depends also on the unknown degree
of turbulence and vertical circulation in gas.

\item Translational lattice modes in water ice appear
in emission at $45\mum$ and possibly also at $62\mum$
in CQ Tau and AA Tau, two of the coolest stars in
our sample ($T_* \lesssim 7200\K$). We interpret
these emission bands as arising from disk atmospheric
silicates mantled by water ice at stellocentric
distances of $\sim$100 AU. The hottest stars
in our sample, MWC 480 and HD 36112 ($T_* \gtrsim 8400\K$),
evince no such emission bands.
By itself, the dependence on stellar
temperature of the location of the ice sublimation boundary in the
disk surface layer is insufficiently steep (approximately
$a_{sub,s} \propto T_*^3$) to account for the presence
and absence, respectively, of water ice bands in LWS spectra
of CQ Tau and HD 36112; these two stars differ in their
effective temperatures by only $\sim$15\%.

\end{enumerate}

\acknowledgments
We thank Cornelia J\"{a}ger and Ted Roush
for providing optical constants of olivine and iron, respectively;
Bruce Draine for suggesting that thermal spiking
of PAHs might be responsible for
observed near-infrared excesses;
Jeroen Bouwman and Rens Waters for sending preprints
of their work; Willem Schutte and Peter Goldreich for
helpful discussions; and Antonella Natta for a useful
referee's report. Support for EIC and MKJ was provided
by NASA grant NAG5-7008.
Additional support for EIC was provided by NASA through a
Hubble Fellowship grant
awarded by the Space Telescope
Science Institute, which is operated by the Association
of Universities for Research in Astronomy, Inc., for
NASA under contract NAS 5-26555. MCE and GAB acknowledge
support through the NASA Origins and ISO block grant programs.
JEK is supported in part by a NASA GSRP fellowship.
EFvD acknowledges NWO grant 614.41.003.

\begin{appendix}
\section{Solution for $T_i(a)$ and $\gamma (a)$}
Equations (\ref{ieb}), (\ref{alpha1}), and (\ref{ha})
combine to yield

\begin{eqnarray}
\label{master}
\sin \left[ {\arctan \left( \gamma \frac{H}{h} \sqrt \frac{T_i}{T_c} \sqrt
\frac{a}{R_*} \right) - \arctan \left( \frac{H}{h} \sqrt \frac{T_i}{T_c} \sqrt
\frac{a}{R_*} \right) + \arcsin \left( \frac{4}{3\pi}\frac{R_*}{a} \right)}
\right] \\ \nonumber
= \frac{2}{\phi} \left(\frac{1- e^{-\Sigma\langle\kappa_i\rangle _i}}{1-
e^{-\Sigma\langle\kappa_i\rangle _s}}\right) \, \left( \frac{T_i}{T_*}
\right)^4 \, \left({\frac{a}{R_*}}\right)^2 \, .
\end{eqnarray}

\noindent This is an equation for $T_i(a)$, where

\begin{equation}
\label{submaster}
\gamma (a) \equiv \frac{d\ln H}{d\ln a} = \frac{3}{2} + \frac{1}{2}\frac{d\ln
T_i}{d\ln a} \, .
\end{equation}

\ni Note that $\langle \kappa_i \rangle_i$ is the interior
opacity averaged over the Planck function evaluated at $T_i$;
$\phi$, $H/h$, $T_c$, $T_*$, and $R_*$ are constants.
We rely on the slow and modest variation of $\gamma$ with distance
to solve for $T_i(a)$ in the following manner. We define a
logarithmic grid in $a$ = $\{$$a_1$, ...~, $a_N$$\}$ where
typically $N = 300$. We begin by guessing a value for
$\gamma$ at $a = a_1$. This value of $\gamma$ is used in
equation (\ref{master}) to solve for {\it both} $T_i(a_1)$ and
$T_i(a_2)$ by Brent's root finder (Press et al.~1992).
These latter values furnish a new $\gamma \rightarrow \gamma ' = 3/2 + (1/2)
\ln [T_i(a_2)/T_i(a_1)] / \ln (a_2/a_1)$ by
(\ref{submaster}). This new value of $\gamma '$ is then
employed in (\ref{master}) to compute $T_i(a_3)$ and $T_i(a_4)$.
These, in turn, furnish $\gamma ''$ for $T_i(a_5)$ and $T_i(a_6)$.
Thus the iteration proceeds by updating $\gamma$
after every two steps in distance.
This procedure quickly converges to a smoothly varying solution
after the first one or two iterations of $\gamma$.
The initial guess of $\gamma$ at $a_1$ can
be improved {\it a posteriori} and the calculation repeated.
Updating $\gamma$ after every one
step in distance introduces numerical instability.
That is, if we take $\gamma '$ to compute
$T_i(a_3)|_{\gamma '}$, then employ $T_i(a_2)|_{\gamma}$
and $T_i(a_3)|_{\gamma '}$ to calculate $\gamma ''$,
and so on, the resultant solution jumps erratically
with every step.

\end{appendix}





\begin{references}
\reference{a00} Allen, C.W. 2000, Allen's Astrophysical Quantities, ed. Arthur
N. Cox (New York : AIP Press)
\reference{b00} Beckwith, S.V.W., Henning, Th., \& Nakagawa, Y. 2000, in
Protostars and Planets IV, eds. V. Mannings, A.P. Boss, \& S.S. Russell (Tucson
: University of Arizona Press), 533--558
\reference{bs91} Beckwith, S.V.W., \& Sargent, A.I. 1991, \apj, 381, 250
\reference{betal90} Beckwith, S.V.W., Sargent, A.I., Chini, R.S., \&
G\"{u}sten, R. 1990, \aj, 99, 924
\reference{bl94} Bell, K.R. \& Lin, D.N.C. 1994, \apj, 427, 987
\reference{blw69} Bertie, J.E., Labb\'{e}, H.J., \& Whalley, E. 1969, J. Chem.
Phys., 50, 4501
\reference{bh83} Bohren, C.F. \& Huffman, D.R. 1983, Absorption and Scattering
of Light by Small Particles (New York : Wiley-Interscience)
\reference{betal00} Bouwman, J., de Koter, A., van den Ancker, M.E., \& Waters,
L.B.F.M. 2000, \aap, 360, 213
\reference{cetal91} Calvet, N., Pati\~{n}o, A., Magris, G., \& D'Alessio, P.
1991, \apj, 380, 617
\reference{cetal92} Calvet, N., Pati\~{n}o, A., Magris, G., \& D'Alessio, P.
1992, Rev. Mexicana Astron. Astrof., 24, 27
\reference{chs00} Calvet, N., Hartmann, L., \& Strom, S.E. 2000, in Protostars
and Planets IV, eds. V. Mannings, A.P. Boss, \& S.S. Russell (Tucson :
University of Arizona Press), 377--399
\reference{cg97} Chiang, E.I. \& Goldreich, P. 1997, \apj, 490, 368 (CG97)
\reference{cg99} Chiang, E.I. \& Goldreich, P. 1999, \apj, 519, 279 (CG99)
\reference{cetal00} Creech-Eakman, M.J., Chiang, E.I., van Dishoeck, E.F., \&
Blake, G.A. 2000, \aap, submitted
\reference{detal98} D'Alessio, P., Cant\'{o}, J., Calvet, N., \& Lizano, S.
1998, \apj, 500, 411
\reference{detal99} D'Alessio, P., Calvet, N., Hartmann, L., Lizano, S., \&
Cant\'{o}, J. 1999, \apj, 527, 893
\reference{d95} Draine, B.T. 1995, in The Physics of the Interstellar
Medium and Intergalactic Medium, ASP Conference Series, 80, 133
\reference{detal96} Dutrey, A., Guilloteau, S., Duvert, G., Prato, L., Simon,
M., Schuster, K., \& M\'{e}nard, F. 1996, \aap, 309, 493
\reference{j94} J\"{a}ger, C., Mutschke, H., Begemann, B., Dorschner, J., \&
Henning, Th. 1994, \aap, 292, 641
\reference{j98} J\"{a}ger, C., Molster, F.J., Dorschner, J., Henning, Th.,
Mutschke, H., \& Waters, L.B.F.M. 1998, \aap, 339, 904
\reference{kh87} Kenyon, S.J. \& Hartmann, L. 1987, \apj, 323, 714
\reference{h93} Hudgins, D.M., Sandford, S.A., Allamandola, L.J., \& Tielens,
A.G.G.M. 1993, ApJS, 86, 713
\reference{iras} Infrared Astronomical Satellite (IRAS) Point
Source Catalog 1988, Joint IRAS Science Working Group
(NASA reference publication: 1190)
\reference{l87} Lada, C. J. 1987, in IAU Symp.~115: Star Forming Regions,
eds.~M.~Peimbert \& J.~Jugaku (Dordrecht : Reidel), 1
\reference{lw84} Lada, C. J. \& Wilking, B. A. 1984, \apj, 287, 610
\reference{mb91} Malbet, F. \& Bertout, C. 1991, \apj, 383, 814
\reference{ms97} Mannings, V. \& Sargent, A.I. 1997, \apj, 490, 792
\reference{msd88} McSween, H.Y., Sears, D.W.G., \& Dodd, R.T. 1988,
in Meteorites and the Early Solar System, eds.~J.F.~Kerridge \& M.S.~Matthews
(Tucson : University of Arizona Press), 114--144
\reference{mjc97} Meyer, D.M., Jura, M., \& Cardelli, J.A. 1998, \apj, 493, 222
\reference{metal99} Miroshnichenko, A., Ivezi\'{c}, Z., Vinkovi\'{c}, D., \&
Elitzur, M. 1999, \apjl, 520, L115 (MIVE99)
\reference{n93} Natta, A. 1993, \apj, 412, 761
\reference{npk93} Natta, A., Prusti, T., \& Krugel, E. 1993, \aap, 275, 527
\reference{nw00} Natta, A., \& Whitney, B.A. 2000, \aap, in press.
\reference{petal94} Pollack, J.B., et al.~1994, \apj, 421, 615
\reference{press92} Press, W. H., Teukolsky, S. A., Vetterling, W. T., \&
Flannery, B. P. 1992, Numerical Recipes in Fortran (Cambridge : Cambridge
University Press)
\reference{q00} Qi, C. 2000, Ph.D. Thesis, California Institute of Technology
\reference{setal99} Sitko, M.L., Grady, C.A., Lynch, D.K., Russell, R.W., \&
Hanner, M.S. 1999, \apj, 510, 408
\reference{tetal00} Thi, W.-F., et al. 2000, \apj, in preparation
\reference{w84} Warren, S.G. 1984, Applied Optics, 23, 1206
\reference{wetal99} Webb, R.A., et al.~1999, \apjl, 512, L63
\end{references}
\end{document}